\documentclass[hidelinks,onefignum,onetabnum]{siamart250106}

\usepackage{lipsum}
\usepackage{amsfonts}
\usepackage{graphicx}
\usepackage{epstopdf}
\usepackage{algorithmic}
\usepackage{color}
\usepackage{xcolor}
\ifpdf
  \DeclareGraphicsExtensions{.eps,.pdf,.png,.jpg}
\else
  \DeclareGraphicsExtensions{.eps}
\fi

\newsiamremark{remark}{Remark}
\newsiamremark{hypothesis}{Hypothesis}
\crefname{hypothesis}{Hypothesis}{Hypotheses}
\newsiamthm{claim}{Claim}

\headers{AWB scheme for 2D RSWE in sphere}{A. González, M.J. Castro and J. Macías}

\title{Asymptotically well-balanced geostrophic reconstruction finite volumes numerical schemes for the 2D rotating NLSWE in spherical coordinates\thanks{Submitted to the editors DATE.
\funding{This work was funded by Funded by ChEESE-2P (EU-EuroHPC JU-101093038) and DT-GEO project (HORIZON-INFRA-2021-TECH-01-01, number 101058129). This research has also been partially funded by MCIN/AEI/10.13039/501100011033 and
 by the “European Union NextGenerationEU/PRTR” through the Grant PDC2022‐133663‐C21 and by MCIN/AEI/10.13039/50110001103 and by “ERDF A way of making Europe,” by the European Union through the Grant PID2022‐137637NB‐C21.}}}

\author{A. González \thanks{Departamento de Análisis Matemático, Estadística e Investigación Operativa y Matemática Aplicada,
Facultad de Ciencias, Universidad de Málaga, Campus de Teatinos, 29010 Málaga, Spain  (\email{alexgp@uma.es}, \email{mjcastro@uma.es}, \email{jmacias@uma.es}).} \and M.J. Castro \footnotemark[2] \and J. Macías \footnotemark[2]}

\usepackage{amsopn}

\makeatletter
\newcommand*{\addFileDependency}[1]{
  \typeout{(#1)}
  \@addtofilelist{#1}
  \IfFileExists{#1}{}{\typeout{No file #1.}}
}
\makeatother

\ifpdf
\hypersetup{
  pdftitle={Asymptotically well-balanced geostrophic reconstruction finite volumes numerical schemes for the 2D rotating NLSWE in spherical coordinates},
  pdfauthor={A. González, M. J. Castro, J. Macías}
}
\fi

\begin{document}

\maketitle

\begin{abstract}
  The dynamics of large-scale geophysical fluids is primarily governed by the balance between the Coriolis force and the pressure gradient. 
  This phenomenon, known as geostrophic equilibrium, is the basis for the geostrophic model, which has proven to be extremely useful for understanding and forecasting large-scale atmospheric and oceanic dynamics. 
  In the present work, we develop second- and third-order finite-volume numerical schemes applied to the 2D rotating shallow-water equations in spherical coordinates. These schemes are designed to preserve the geostrophic equilibrium in the limit as the Rossby number tends to zero. 
  The final goal is to design reliable and efficient forecasting models for simulating meteotsunamis, long-wave events generated in the ocean by atmospheric pressure disturbances. These disturbances produce long waves of small amplitude that gradually amplify as they approach the coast. 
  The numerical results for various analytical and real-world test cases underscore the importance of maintaining geostrophic equilibrium over time.
\end{abstract}

\begin{keywords}
  rotating shallow water equations, finite volumes, geostrophic equilibrium, GPU, meteotsunamis, asymptotic well-balanced
\end{keywords}

\begin{AMS}
  86A15, 65L05, 65M99, 65M08, 65-04

\end{AMS}

\section{Introduction} \label{sec:intro}
Understanding atmospheric and oceanic dynamics has long captured the interest of scientists in many disciplines. All models aimed at addressing these problems use the Navier-Stokes equations as starting point, as they provide the complete set of equations describing fluid behavior. 
However, Navier-Stokes (NS) system of partial differential equations is too overly costly to directly solve them in big domains such as ocean basins or at global scale.
To overcome this challenge, simplifying assumptions based on the specific problem are made, focusing on the dominant dynamics and neglecting less significant terms. This is where the Non-Linear Shallow Water Equations (NLSWE) come into play.
The 2D NLSWE is a system of PDEs derived from the 3D Navier-Stokes system under the assumption that the horizontal scale far exceeds the vertical scale.
As a result, horizontal dynamics dominate fluid movement, whereas vertical influences are comparatively negligible. 
In  addition, the fluid pressure is assumed to be hydrostatic and the horizontal viscosity terms are neglected. 
To further simplify, the horizontal velocity field is depth-averaged, eliminating vertical dependence on the fluid velocities in the Navier-Stokes equations.
The NLSWE are widely used and have demonstrated excellent accuracy in many applications \cite{Zeitlin2018, GarciaNavarro2019}. 
However, when addressing large-scale geophysical flows, two critical factors must be considered: (1) the necessity of a spherical coordinate system, as the geometric terms associated to the Earth's curvature cannot be neglected; and (2) a rotational framework, where the forces due to the Earth's rotation may be retained as they play a significant role. 
When incorporating Earth's rotational effects, a widely used model is the 2D rotating nonlinear shallow-water equations with variable bottom topography (\ref{2dRtotatingSWE}) (see \cite{Pedlosky1987} pp. 59-63 for a full derivation):
\begin{equation}  \label{2dRtotatingSWE}
\begin{cases}
    \partial_{t} h+\nabla \cdot (h\vec{u})=0 \\
    \partial_{t} \vec{u}+\vec{u}\cdot \nabla \vec{u}+\frac{1}{Fr^2}\nabla h + \frac{1}{Ro}2e_{z}\times \vec{u}=\frac{1}{Fr^2}\nabla H 
\end{cases}
\end{equation}
\noindent where $h$ is the water thickness; $\vec{u}$, the velocity vector; $e_z$, the normal unit vector in the vertical direction; $H$, the bottom topography measured from a fixed level of reference; $Fr=\frac{U}{\sqrt{gB}}$, the Froude number; $Ro=\frac{U}{fL}$, the Rossby number; $g$, the gravity acceleration and $U$, $B$, $L$ are typical velocity, height, and length, respectively, and finally $f$ is the Coriolis parameter. Note that the previous formulation is only valid for smooth solutions. If shocks appear in finite time, conserved variables should be used. In this work we will use the conservative formulation of system \eqref{2dRtotatingSWE}

The Rossby number, a dimensionless quantity defined as $Ro=\frac{U}{fL}$, is an important parameter in determining the regime of these fluids. 
When fluid simulations involve a time scale $T=\frac{L}{U}$ that is significantly longer than the Earth's rotation time scale (or equivalently, greater than the inverse of the Coriolis parameter, $\frac{1}{f}$), the system operates in a low Rossby number regime, since the Rossby number can also be expressed as $Ro=\frac{1}{fT}.$

\noindent Under these conditions, the Coriolis force dominates over the non-linear advective terms. 
As a result, the horizontal momentum equations of the fluid reach a balance between the Coriolis force and the pressure gradient \cite{Phillips1963, Blumen1972}. 
This state, where the Coriolis force exactly balances the pressure gradient, is known as \textbf{geostrophic equilibrium}.

The theory of geostrophic equilibrium is a powerful tool for understanding the dynamics of atmospheric and oceanic currents. 
For example, it explains why hurricanes or storms take on spiral shapes and accounts for the ``Taylor Column'' structure observed in rapidly rotating homogeneous incompressible fluids with flat bottom \cite{Hide1971, Sun2001}. 
In Cartesian coordinates, geostrophic equilibrium can be written as:
\begin{equation} \label{cartesian_geostrophic}
\begin{cases}
$$g \, \partial_{x} (h-H) = fv, \\
g \, \partial_{y} (h-H) = -fu, $$
\end{cases}
\end{equation}
where $(u,v)$ is the horizontal velocity field.
Although this theory provides an excellent model for understanding how large-scale fluids flow from high-pressure to low-pressure regions -essentially driven by the geostrophic relation (\ref{cartesian_geostrophic}), which indicates that the movement is parallel to the isobars-, it has several important limitations. 
First, it describes a steady-state equilibrium, making it unsuitable for capturing time-evolving dynamics. 
Additionally, the assumption of nearly zero Rossby number may not hold throughout the system's evolution, meaning that the effects of the non-linear advective terms in (\ref{2dRtotatingSWE}) should be considered during certain phases of the flow. 
To enhance the model, small perturbations around the geostrophic equilibrium are introduced, distinguishing between the velocities associated with the geostrophic and the non-geostrophic components, as their scales differ. 
This new consideration, along with the assumption of a low Froude and Rossby number of the same order of magnitude, leads to the quasi-geostrophic model.
In this framework, the time dependence of the advective terms explicitly appears, and their influence in the fluid's overall behavior is determined by the magnitude of the Rossby number, which is still assumed to be small $Ro\ll 1$ (for a full derivation of the quasi-geostrophic model, see Chapter 4 of \cite{Majda2012}). 

The 2D rotating NLSWE can be interpreted as an approximation of the quasi-geostrophic model in regimes with low Rossby and Froude numbers \cite{Majda2012}. In these regimes, geostrophic balance plays a critical role. As a result, numerical schemes designed to solve these equations often focus on preserving an exact or highly accurate solution of the geostrophic equilibrium.
In other words, these schemes are well-balanced for such steady states. 
Well-balanced schemes for a specific steady state -or a set of steady states- enable the accurate capture of small perturbations over the stationary state, even on coarse meshes, as the stationary state is resolved to machine precision (see \cite{Arpaia2022, BOUCHUT2004, Castro2017, Castro2020, CastroDiaz2018, Chertock2017, Desveaux2021, Gallardo2008, GonzalezTabernero2024, GomezBueno2021, LukacovaMedvid’ova2007, NavasMontilla2018} and references therein).

It is important to note that the geostrophic equilibrium is \textit{not} a steady state of the 2D rotating NLSWE, although it is for the linearized rotating shallow water equations (SWE). Finding a general steady state for the rotating SWE requires determining a non-zero velocity field, but research on this topic remains limited, even in Cartesian coordinates. 
A specific family of geostrophic equilibria, known as \textit{jets in a rotational frame}, are solutions to (\ref{cartesian_geostrophic}), where the fluid moves along streamlines at constant velocity. 
These solutions are expressed as follows:
\begin{align}
&u=0, \quad \partial_y v = 0, \quad \partial_y h=0, \quad \partial_y \eta =0, \quad g\partial_x \eta = fv, \\
&v=0, \quad \partial_x u = 0, \quad \partial_x h=0, \quad \partial_x \eta=0, \quad g\partial_y \eta = -fu, \\
&u,v \neq 0, \, \, v=ku, \quad g\partial_x \eta = fv, \quad g\partial_y \eta=-fu, \quad k=tan(\theta)=\mbox{constant}
\end{align}
\noindent where $\eta=h-H$ is the water-free surface; $\theta$, the rotation angle of the streamline direction. 

A thorough review of the existing literature reveals a variety of strategies for designing numerical schemes aimed at preserving the geostrophic equilibrium for the rotating NLSWE in 1D and 2D on Cartesian grids. \cite{BOUCHUT2004, LukacovaMedvid’ova2007, Castro2008, Audusse2011, Audusse2018, Chertock2017, NavasMontilla2018, Audusse2021, Desveaux2021, Do2023, GonzalezTabernero2024}. 
For the 1D case, Bouchut et al. \cite{BOUCHUT2004} designed a second-order numerical scheme that preserves the steady state of the water-at-rest. 
In their work, they introduced a geometric interpretation of the Coriolis source term, which has significantly influenced subsequent research on this particular topic. 
This reinterpretation, known as the ``apparent topography'' method, addresses the well-balanced problem by considering the primitive variables for the Coriolis terms, with the aim of preserving them. 
Castro et al. \cite{Castro2008} focused on simulating the geostrophic adjustment phenomenon by proposing second- and third-order schemes with a special time discretization to better approximate inertial oscillations. 
Desveaux et al. \cite{Desveaux2021} developed fully well-balanced first- and second-order methods based on Godunov-type schemes. 
Gonz\'alez Tabernero et al. \cite{GonzalezTabernero2024} proposed a general procedure for constructing arbitrary high-order fully well-balanced schemes, applying them to produce first-, second-, and third-order methods.

In the case of the 2D system, Luká$\check{c}$ová et al. \cite{LukacovaMedvid’ova2007} implemented first- and second-order finite volume evolution Galerking methods that are able to preserve jets in rotational frame. 
Audusse et al. \cite{Audusse2011} developed a well-balanced first-order scheme by reinterpreting the discretization of the Coriolis source term and introducing an auxiliary pressure that is a solution of a Laplace equation using a dual grid, obtaining remarkable results even for high Froude numbers. 
This paper was continued by several publications: in \cite{Audusse2018}, the authors analyze why classical Godunov-type schemes fail when dealing with flows close to the geostrophic equilibrium, and they develop a collocated Godunov finite volume scheme for the linear version of the RSWE modifying diffusive terms in the classical scheme; 
in \cite{Audusse2021}, new finite volume numerical schemes for collocated and staggered meshes were developed to improve upon the standard methods. They highlighted the good behavior of the linear versions of these schemes near a geostrophic state. Additionally, the study emphasizes the importance of accurately approximating energy conservation to ensure the stability of the methods in the nonlinear case.
Navas-Montilla et al. \cite{NavasMontilla2018} propose an arbitrary order numerical scheme for the non-rotating and rotating NLSWE that is well-balanced for both the water-at-rest and the jet in rotational frame. Their approach is based on an extension to the 2D case of a previous 1D ARL-ADER scheme. While their results demonstrate high accuracy, they note that this accuracy comes with a significant computational cost.
Chertock et al. \cite{Chertock2017} implement a second-order well-balanced central-upwind and finite volume evolution Galerking scheme that maintain the jet in rotational frame. They aim at reconstructing the Coriolis primitive variables to achieve the well-balanced property. Their scheme works well for situations close to the geostrophic equilibrium.

Research in the spherical framework of the NLSWE has also made significant progress \cite{Bernard2009, Bonev2018, Castro2017, Arpaia2022}. 
Castro et al. \cite{Castro2017} introduced a well-balanced high-order path-conservative finite volume numerical scheme for the non-rotating NLSWE. Their method preserves the water-at-rest solutions and effectively addresses the singularity at the poles. 
Bonev et al. {\cite{Bonev2018} derived a novel high-order Discontinuous Galerkin (DG) scheme capable of preserving water-at-rest solutions and with a robust treatment of the wet/dry fronts. 
Arpaia et al. \cite{Arpaia2022} extended the work of \cite{Bernard2009} by developing a mixed 3D/2D covariant DG method. Their scheme achieves the well-balanced property for the inverted barometer state, which occurs when hydrostatic pressure, bottom forces, and atmospheric pressure are in equilibrium.

None of the aforementioned studies have attempted to preserve or approximate the geostrophic balance using finite volume schemes within the framework of spherical coordinates. To our knowledge, this remains an unexplored area. The primary objective of our work is to develop a family of high-order, efficient, well-balanced finite volume methods for the 2D rotating NLSWE on the sphere. These methods aim to preserve geostrophic equilibrium effectively in this context. The proposed numerical schemes will be suitable for accurately simulating trans-oceanic tsunami-like waves, where both the Coriolis effect and Earth's curvature play an crucial role. Additionally, the numerical schemes are designed to model meteotsunami phenomena -an emerging area of research in natural hazards that has garnered increasing interest among the tsunami community, as evidenced by the growing number of related publications \cite{Monserrat2006, Vilibic2008, Denamiel2019, Denamiel2020}.
Notably, the Hunga Tonga-Hunga Ha’apai event in 2022 has highlighted the significance of this phenomenon \cite{Amores2022, Anup2024, Boslough2024, Denamiel2023, Purkis2023}.
Meteotsunamis are generated by atmospheric disturbances with steep gradients of pressure and/or wind. 
The most extreme meteotsunami events are driven by two primary mechanisms: offshore amplification of oceanic long-waves due to Proudman or Greenspan resonances (where the atmospheric disturbance travels at the same speed as the oceanic long-waves), and nearshore amplification caused by resonance frequencies associated with the geometry of shelves, bays, or inlets.

This work extends the approach presented in \cite{Castro2017} to the rotating shallow water system, incorporating spatially and temporally varying atmospheric pressure. 
The structure of the paper is as follows. 
In Section \ref{sec:2}, the PDE system is introduced and a dimensionless version is derived. This formulation facilitates the inclusion of additional pressure terms without introducing unnecessary complexity. 
In Section \ref{sec:3}, the semidiscrete numerical scheme in space is presented, introducing the path-conservative procedure and the reconstruction operator used to achieve high-order in space. 
Section \ref{sec:4} is devoted to the geostrophic equilibrium and the well-balanced properties of the numerical scheme. 
It begins with a derivation of the geostrophic balance in spherical coordinates via linearization. Then, second- and third-order geostrophic reconstructions are prescribed. 
The asymptotic behavior of the well-balanced numerical procedure to the geostrophic equilibrium is then demonstrated.
Section \ref{sec:5} introduces numerical tests to evaluate the performance of the proposed schemes.
Finally, the paper concludes with a summary of the work presented and some final remarks.
    
\section{PDE system} \label{sec:2}

The shallow water system on the sphere, accounting for bottom topography, the Coriolis force, and non-constant, varying in time and space, atmospheric source terms, can be expressed as a system of balance laws \cite{Tort2014, Castro2017} of the form:
\begin{equation} \label{2DRotatingSWESphere}
\begin{cases}
$$\partial_{t}h_{\sigma}+\frac{1}{R}\bigg[\partial_{\theta}\bigg(\frac{Q_{\theta}}{\sigma}\bigg)+\partial_{\varphi}Q_{\varphi}\bigg]=0, \\

\partial_{t}Q_{\theta}+\frac{1}{R}\partial_{\theta}\bigg(\frac{Q_{\theta}^2}{h_{\sigma}\sigma}\bigg)+\frac{1}{R}\partial_{\varphi}\bigg(\frac{Q_{\theta}Q_{\varphi}}{h_{\sigma}}\bigg)+\frac{Q_{\theta}Q_{\varphi}}{Rh_{\sigma}\sigma}\partial_{\varphi}\sigma\\
\quad \quad \quad \quad \quad \quad \quad \quad +\frac{gh_{\sigma}}{R\sigma^2}\partial_{\theta}\eta_{\sigma}=fQ_{\varphi}-\frac{h_{\sigma}}{\rho R \sigma}\partial_{\theta}p^{a}, \\

\partial_{t}Q_{\varphi}+\frac{1}{R}\partial_{\theta}\bigg(\frac{Q_{\theta}Q_{\varphi}}{h_{\sigma}\sigma}\bigg)+\frac{1}{R}\partial_{\varphi}\bigg(\frac{Q_{\varphi}^2}{h_{\sigma}}\bigg)-\bigg(\frac{Q_{\theta}^2}{Rh_{\sigma}\sigma}+\frac{gh_{\sigma}\eta_{\sigma}}{R\sigma^2}\bigg)\partial_{\varphi}\sigma\\
\quad \quad \quad \quad \quad \quad \quad \quad +\frac{gh_{\sigma}}{R\sigma}\partial_{\varphi}\eta_{\sigma}=-fQ_{\theta}-\frac{h_{\sigma}}{\rho R}\partial_{\varphi}p^{a}, $$

\end{cases}
\end{equation}


\noindent where $R$ is the radius of the Earth; $(\theta,\varphi)$ are longitude and latitude coordinates, resp.; $g$, the gravity; $\sigma=\mbox{cos}(\varphi)$; $h_{\sigma}=h\sigma$, being $h$ the thickness of the water layer; $Q_{\theta}=q_{\theta}\sigma$, being $q_{\theta}$ the discharge in the longitude direction; $Q_{\varphi}=q_{\varphi}\sigma$, being $q_{\varphi}$ the discharge in the latitude direction; $\eta_{\sigma}=\eta \sigma$, being $\eta = h - H$ the free surface elevation; $H$, the bottom topography measure from a reference level (usually the mean sea level); $p^{a}$, the time and space dependent atmospheric pressure; $f=2 \, \Omega \, \mbox{sin}(\varphi)$, the latitude-dependent Coriolis parameter; $\Omega$, the Earth's angular velocity; and $\rho$, the water density. 
Note that the spherical approach leads to additional terms multiplied by $\partial_{\varphi}\sigma$, which are intrinsically related to the curvature of the Earth. 

Following a similar approach to that in \cite{Arpaia2022}, the atmospheric pressure can be interpreted as an apparent bottom that changes over time. Consequently, the definition of the water-free surface can be modified as $\widetilde{\eta_{\sigma}}:=\eta_{\sigma}+p_{\sigma}^a=h_{\sigma}-H_{\sigma}+p_{\sigma}^a=h_{\sigma}-(H_{\sigma}-p_{\sigma}^a)= h_{\sigma}-\widetilde{H_{\sigma}}$, where $p_{\sigma}^{a}=p^{a}\sigma$. 
Furthermore, given the characteristic length, $L$, and the characteristic height, $B$, the characteristics velocity and time can be defined as $U=\sqrt{gB}$ and $T=\frac{L}{U}$, respectively. 
With these definitions, the dimensionless form of the system (\ref{2DRotatingSWESphere}) can be derived, where $g=1$, $\rho=1$, allowing the system to be compactly written as:
\begin{equation} \label{2DRotatingSWESphereCompact}
\begin{cases}
\partial_{t}h_{\sigma}+\frac{1}{R}\bigg[\partial_{\theta}\bigg(\frac{Q_{\theta}}{\sigma}\bigg)+\partial_{\varphi}Q_{\varphi}\bigg]=0, \\

\partial_{t}Q_{\theta}+\frac{1}{R}\partial_{\theta}\bigg(\frac{Q_{\theta}^2}{h_{\sigma}\sigma}\bigg)+\frac{1}{R}\partial_{\varphi}\bigg(\frac{Q_{\theta}Q_{\varphi}}{h_{\sigma}}\bigg)+\frac{Q_{\theta}Q_{\varphi}}{Rh_{\sigma}\sigma}\partial_{\varphi}\sigma+\frac{h_{\sigma}}{R\sigma^2}\partial_{\theta}\widetilde{\eta_{\sigma}}=fQ_{\varphi},\\

\partial_{t}Q_{\varphi}+\frac{1}{R}\partial_{\theta}\bigg(\frac{Q_{\theta}Q_{\varphi}}{h_{\sigma}\sigma}\bigg)+\frac{1}{R}\partial_{\varphi}\bigg(\frac{Q_{\varphi}^2}{h_{\sigma}}\bigg)-\bigg(\frac{Q_{\theta}^2}{Rh_{\sigma}\sigma}+\frac{h_{\sigma}\widetilde{\eta_{\sigma}}}{R\sigma^2}\bigg)\partial_{\varphi}\sigma\\
\quad \quad \quad \quad \quad \quad \quad \quad \quad \quad \quad \quad \quad \quad \quad \quad \quad \quad \quad \quad \quad +\frac{h_{\sigma}}{R\sigma}\partial_{\varphi}\widetilde{\eta_{\sigma}}=-fQ_{\theta}.
\end{cases}
\end{equation}
In this formulation, the terms associated with the atmospheric pressure are integrated into the modified definition of the water-free surface. 
For simplicity, the variables in system (\ref{2DRotatingSWESphereCompact}) retain the same names as those in (\ref{2DRotatingSWESphere}), although they now represent dimensionless quantities. 
The dimensionless system (\ref{2DRotatingSWESphereCompact}) can be expressed in vector form as:
\begin{equation} \label{2DRotatingSWESphereCompactVectorSolvedSystem}
    \partial_{t}w+\frac{1}{R}\big[\partial_{\theta}F_{\theta}(W)+\partial_{\varphi}F_{\varphi}(W)+T_{\theta}^{p}(W)\partial_{\theta}\widetilde{\eta_{\sigma}}+T_{\varphi}^{p}(W)\partial_{\varphi}\widetilde{\eta_{\sigma}}+G_{\varphi}(W)\partial_{\varphi}\sigma\big]=0,
\end{equation}
being
\begin{align*}
&W=\left( \begin{array}{c} w \\ \sigma \end{array} \right)=\left( \begin{array}{c} h_{\sigma} \\ Q_{\theta} \\ Q_{\varphi} \\ \sigma \end{array} \right), \quad F_{\theta}(W)=\left( \begin{array}{c} \frac{Q_{\theta}}{\sigma} \\ \frac{Q_{\theta}^2}{h_{\sigma}\sigma} \\ \frac{Q_{\theta}Q_{\varphi}}{h_{\sigma}\sigma} \end{array} \right), \\
&F_{\varphi}(W)=\left( \begin{array}{c} Q_{\varphi} \\ \frac{Q_{\theta}Q_{\varphi}}{h_{\sigma}} \\ \frac{Q_{\varphi}^2}{h_{\sigma}} \end{array} \right), \quad T_{\theta}^{p}(W)=\left( \begin{array}{c} 0 \\ \frac{h_{\sigma}}{\sigma^2} \\ 0 \end{array} \right),\\
&T_{\varphi}^{p}(W)=\left( \begin{array}{c} 0 \\ 0 \\ \frac{h_{\sigma}}{\sigma} \end{array} \right), \quad G_{\varphi}(W)=G_{\varphi}^{1}(W)+G_{\varphi}^{2}(h_{\sigma},\widetilde{\eta_{\sigma}},\sigma),\\
&G_{\varphi}^{1}(W)=\left( \begin{array}{c} 0 \\ \frac{Q_{\theta}Q_{\varphi}}{h_{\sigma}\sigma}+2\Omega Q_{\varphi}R \\ -\frac{Q_{\theta}^2}{h_{\sigma}\sigma}-2\Omega Q_{\theta}R \end{array} \right), \quad G_{\varphi}^{2}(h_{\sigma},\widetilde{\eta_{\sigma}},\sigma)=\left( \begin{array}{c} 0 \\ 0 \\ -\frac{h_{\sigma}\widetilde{\eta_{\sigma}}}{\sigma^2} \end{array} \right).\\
&
\end{align*}
We are dealing with a hyperbolic system that includes non-conservative products and geometric source terms. 
More precisely, the components of this system are: $\vec{F}=(F_{\theta},F_{\varphi})$, the advective flux; $\vec{T}^{p}=(T_{\theta}^{p},T_{\varphi}^{p})$, the pressure term which gives rise to the non-conservative product $\vec{T^{p}} \cdot \nabla \widetilde{\eta_{\sigma}}$; $G_{\varphi}(W)\partial_{\varphi} \sigma$, the source term. 
Note that the Coriolis parameter can be written as $f=-2\, \Omega \partial_{\varphi} \sigma$, allowing it to be included within the geometric source term. Furthermore, both the advective flux and the pressure term satisfy the rotational invariance-like properties:
\begin{equation} \label{rotationalInvariaceLike}
R_{\vec{\nu}}F_{\vec{n}}(W)=\delta F(R_{\vec{\nu}}w) 
\quad \hbox{and} \quad
R_{\vec{\nu}}T_{\vec{n}}^{p}(W)=\delta T^{p}\big(\tfrac{h_{\sigma}}{\sigma}\big),
\end{equation}
\noindent where
\begin{equation} \label{NormalFlux}
F_{\vec{n}}(W)=n_{\theta}F_{\theta}(W)+n_{\varphi}F_{\varphi}(W), \\
\quad T_{\vec{n}}^{p}(W) = n_{\theta}T_{\theta}^{p}(W)+n_{\varphi}T_{\varphi}^{p}(W),
\end{equation}
%
%
%
%
%
%
\begin{equation} \label{RotationStuff}
\begin{split}
&\delta=\sqrt{\bigg(\frac{n_{\theta}}{\sigma} \bigg)^2+(n_{\varphi})^{2}}\, \,, \quad \vec{\nu}=\left( \begin{array}{c} \frac{n_{\theta}}{\sigma \delta} \\ \frac{n_{\varphi}}{\delta} \end{array} \right), \\
&\vec{\nu}^{\perp}=\left( \begin{array}{c} -\nu_{\varphi} \\  \nu_{\theta}\end{array} \right), \quad R_{\vec{\nu}}=\left( \begin{array}{ccc} 1 & 0 & 0 \\ 0 & \nu_{\theta} & \nu_{\varphi} \\ 0 & -\nu_{\varphi} & \nu_{\theta} \end{array} \right), 
\end{split}
\end{equation}
\noindent and for every $U_{\sigma}=[h_{\sigma},Q_{\vec{\nu}},Q_{\vec{\nu}^{\perp}}]^{T}$ and $h$:
%
%
\begin{equation} \label{1dAssociatedFunctions}
  F(U_{\sigma})=\left( Q_{\vec{\nu}}, \frac{Q_{\vec{\nu}}^2}{h_{\sigma}}, \frac{Q_{\vec{\nu}}Q_{\vec{\nu}^{\perp}}}{h_\sigma} \right)^{T}, \quad T^{p}(h)=\left(0, h, 0 \right)^{T}.  
\end{equation}

\section{Numerical Scheme} \label{sec:3}

The semidiscrete high-order in space numerical method implemented in this work is based on a first-order path-conservative numerical scheme \cite{Castro2009, CastroDiaz2012, Castro2013} and reconstruction operators \cite{Leer1979, Harten1997, Gallardo2011, Cravero2015, Cravero2017, CastroDiaz2018} following the procedure detailed in \cite{Castro2017}.

To provide the expression of a semidiscrete high-order in space path-conservative method for (\ref{2DRotatingSWESphereCompactVectorSolvedSystem}), the computational domain $D = [-\pi,\pi]\times (-\frac{\pi}{2},\frac{\pi}{2})$ in the $\theta \, - \, \varphi$ plane is partitioned into volumes $\mathcal{T} = \{V_i\}_{i=1}^N$. {The set of volumes $\mathcal{T}$ can be either a structured or an unstructured partition of the domain, with the most common cases being triangulated (unstructured) or regular Cartesian (structured) grids. In many geophysical applications related to tsunamis or meteotsunamis, using structured rectangular grids in longitude-latitude coordinates is often advantageous, since both topobathymetric data and other relevant databases are usually stored on regular Cartesian grids. Furthermore, this regular grid structure enhances computational efficiency, particularly on GPUs, due to the ease of memory access that structured data layouts enable.
%

For this reason, and for simplicity, we chose to employ a regular Cartesian grid, where each volume is denoted as $V_{i}=[\theta_{i-\frac{1}{2}},\theta_{i+\frac{1}{2}}]\times[\varphi_{i-\frac{1}{2}},\varphi_{i+\frac{1}{2}}]$ and each point as $\textbf{x} = (\theta, \varphi)$.} Each pair $V_i$, $V_j$ either shares a common edge $E_{i,j}$, a common vertex or their intersection is empty. 
In the case of a common vertex, $E_{i,j}$, the vector $\vec{n}_{i,j}=[n_{i,j}^{\theta},n_{i,j}^{\varphi}]$ represents the normal unit vector to $E_{i,j}$ that points from $V_i$ to $V_j$. 
$|V_i|$ and $|E_{i,j}|$ represent the area of $V_i$ and the length of $E_{i,j}$,  respectively. The set of the indexes of the four neighbor volumes of a given cell $V_i$ that share an edge with it is defined as $\mathcal{N}_i$. 
In a similar way as done in \cite{Castro2017}, we can write the semidiscrete in space numerical scheme (the dependence of $t$ is omitted to simplify the notation) as:
\begin{equation} \label{semidiscreteNumericalScheme}
\begin{split} 
&w_{i}'(t)=-\frac{1}{R|V_{i}|}\bigg[\sum_{j\in\mathcal{N}_{i}}\bigg(\int_{E_{i,j}}F_{\vec{n}_{i,j}}(W_{i,j}^{-}(\gamma)) \, d\gamma+\int_{E_{i,j}}\delta_{i,j}(\gamma)\mathcal{D}_{i,j}^{-}(\gamma)\, d\gamma \bigg)\\
+&\int_{V_i}\big(T_{\theta}^{p}(P_{i}(\textbf{x}))\partial_{\theta}p_{\widetilde{\eta_{\sigma}},i}(\textbf{x})+T_{\varphi}^{p}(P_{i}(\textbf{x}))\partial_{\varphi}p_{\widetilde{\eta_{\sigma}},i}(\textbf{x})\big) \, d\textbf{x}+\int_{V_i}G_{\varphi}(P_i(\textbf{x}))\partial_{\varphi}\sigma(\textbf{x}) \, d\textbf{x}\bigg]
\end{split}
\end{equation}
\noindent where $$W_{i}(t)=\left( \begin{array}{c} w_{i}(t) \\ \bar{\sigma}_{i} \end{array} \right)=\left( \begin{array}{c} h_{\sigma, i}(t) \\ Q_{\theta, i}(t) \\ Q_{\varphi, i}(t) \\ \bar{\sigma}_{i} \end{array} \right)$$ 
is the approximation of the average of the solution in volume $V_i$ at time $t$ ($\bar{\sigma}_{i} $ is the exact average).
$$P_i(\textbf{x})=\left( \begin{array}{c} p_{h_{\sigma},i}(\textbf{x}) \\ p_{Q_{\theta},i}(\textbf{x}) \\ p_{Q_{\varphi},i}(\textbf{x}) \\ \sigma_{i}(\textbf{x}) \end{array} \right), \quad W_{i,j}^{\pm}(\gamma)=\left( \begin{array}{c} w_{i,j}^{\pm}(\gamma) \\ \sigma(\gamma) \end{array} \right)=\left( \begin{array}{c} h_{\sigma,i,j}^{\pm}(\gamma) \\ Q_{\theta,i,j}^{\pm}(\gamma) \\ Q_{\varphi,i,j}^{\pm}(\gamma) \\ \sigma(\gamma) \end{array} \right)$$ are the reconstructions of the averages $\{h_{\sigma,i}\}, \, \{Q_{\theta,i}\}, \, \{Q_{\varphi,i}\}$ at $\textbf{x} \in \mathring{V}_i$ and the reconstructed states at $\gamma \in E_{i,j}$, resp.;  $p_{\widetilde{\eta_{\sigma}},_i}(\textbf{x})$ and $\widetilde{\eta_{\sigma}}_{i,j}^{\pm}(\gamma)$ are the reconstructions of $\{\widetilde{\eta_{\sigma}},_{i}\}$ at $\textbf{x} \in \mathring{V}_i$ and $\gamma \in E_{i,j}$, respectively. $F_{\vec{n}_{i,j}}(W)$ is the normal advective flux (\ref{NormalFlux}) to the edge $E_{i,j}$ and $\mathcal{D}_{i,j}^{-}(\gamma)$ is the incoming fluctuation from the edge $E_{i,j}$, which accounts for the jump in the flux and the pressure terms. It is defined as 
\begin{equation} \label{2DFluctuation}
\mathcal{D}_{i,j}^{-}(\gamma)=R^{-1}_{\vec{\nu}_{i,j}(\gamma)}\cdot\mathcal{D}^{-}(R_{\vec{\nu}_{i,j}(\gamma)}w_{i,j}^{-}(\gamma),\widetilde{\eta_{\sigma}}_{i,j}^{-}; \, R_{\vec{\nu}_{i,j}(\gamma)}w_{i,j}^{+}(\gamma),\widetilde{\eta_{\sigma}}_{i,j}^{+}),
\end{equation}
\noindent being $\mathcal{D}^{-}$ a fluctuation function of a first-order path-conservative method for the 1D system of balance laws described in supplementary materials (SM1). Finally, $\delta_{i,j}(\gamma)$, $\vec{\nu}_{i,j}(\gamma)$ and $R_{\vec{\nu}_{i,j}(\gamma)}$ are the terms from (\ref{RotationStuff}) applied to the normal vector $\vec{n}_{i,j}$ and evaluated at $\gamma \in E_{i,j}$.

In practice, the integrals that appear in (\ref{semidiscreteNumericalScheme}) are approximated using quadrature formulas with an order equal to or greater than that of the reconstruction operators. 
This results in the following general expression:
\begingroup
\small
\begin{equation} \label{semidiscreteNumericalSchemeUsingQuadrature}
\begin{split} 
&w_{i}'(t)=-\frac{1}{R|V_{i}|}\bigg[\sum_{j\in\mathcal{N}_{i}}\bigg(|E_{i,j}|\sum_{l=0}^k \alpha_l F_{\vec{n}_{i,j}}(W_{i,j}^{-}(\gamma_{i,j}^l))+|E_{i,j}|\sum_{l=0}^k \alpha_l \delta_{i,j}(\gamma_{i,j}^l)\mathcal{D}_{i,j}^{-}(\gamma_{i,j}^l)\bigg)\\
&+|V_i|\sum_{l=0}^K \beta_l \bigg(T_{\theta}^{p}(P_{i}(\textbf{x}_{i}^l))\partial_{\theta}p_{\widetilde{\eta_{\sigma}},i}(\textbf{x}_{i}^l)+T_{\varphi}^{p}(P_{i}(\textbf{x}_{i}^l))\partial_{\varphi}p_{\widetilde{\eta_{\sigma}},i}(\textbf{x}_{i}^l)+G_{\varphi}(P_i(\textbf{x}_{i}^l))\partial_{\varphi}\sigma(\textbf{x}_{i}^l)\bigg) \, \bigg],
\end{split}
\end{equation}
\endgroup
%
%
\noindent where $\{\gamma_{i,j}^l \}_{l=0}^k$, $\{\alpha_{l}\}_{l=0}^k$ represent the quadrature points and weights of the formula applied to $E_{i,j}$; while $\{\textbf{x}_{i}^l\}_{l=0}^K$, $\{\beta_{l}\}_{l=0}^K$, are the corresponding points and weights for the quadrature formula applied to $V_i$. 
Additionally, the volume quadrature rule can be also used to compute the initial cell averages of the conserved variables $w_i^{0}$, along with the cell averages $\bar{\sigma}_{i}$.

The time discretization is performed by applying a high-order Total Variation Diminishing Runge-Kutta (TVD RK) time integrator \cite{Gottlieb1998} to the ODE system (\ref{semidiscreteNumericalSchemeUsingQuadrature}). 
It is important to note that the reconstruction operator is applied to $\widetilde{\eta}_{\sigma}$ rather than $\widetilde{H}_{\sigma}$, with the reasons for this choice discussed in Section \ref{sec:4}. 

Two numerical schemes are proposed in this work: (1) a second-order in space scheme based on a MUSCL reconstruction operator \cite{Leer1979} coupled with a slope limiter \cite{Serna2005}, and a second-order TVD RK time integrator, together with the mid-point quadrature rule; and (2) a third-order in space scheme based on a central weighted essentially non-oscillatory (CWENO) reconstruction operator \cite{CastroDiaz2018}, and third-order TVD RK time integrator together with the two-points Gauss quadrature formula.
Equivalently, setting $\textbf{x}_{i,j}$ as the center point on edge $E_{i,j}$, and $\textbf{x}_{i}=(\theta_i, \varphi_i)$ as the center of volume $V_i$, then expression 
 (\ref{semidiscreteNumericalSchemeUsingQuadrature}) for the second-order scheme is determined by the following values $k=0, \quad K=0, \quad \alpha_{0}=1, \quad \gamma_{i,j}^0=\textbf{x}_{i,j}, \quad \textbf{x}_{i}^0=\textbf{x}_i.$
For the third-order scheme, the corresponding values are:
\setlength{\jot}{2pt} 
\begin{align*} 
&k=1, \quad K=3, \quad \alpha_{0}= \alpha_{1}=\tfrac{1}{2}, \quad \beta_{0}= \beta_{1}=\beta_{2}=\beta_{3}=\tfrac{1}{4},\\
&\gamma_{i,j}^0=\textbf{x}_{i,j}+\Delta \varphi \sqrt{1/12}, \quad \gamma_{i,j}^1=\textbf{x}_{i,j}-\Delta \varphi \sqrt{1/12}, \quad \mbox{if} \, E_{i,j} \, \mbox{is a vertical edge},\\
&\gamma_{i,j}^0=\textbf{x}_{i,j}+\Delta \theta \sqrt{1/12}, \quad \,\gamma_{i,j}^1=\textbf{x}_{i,j}-\Delta \theta \sqrt{1/12}, \quad \, \mbox{if} \, E_{i,j} \, \mbox{is an horizontal edge}, \\
&\textbf{x}_{i}^{0}=(\theta_i - \Delta \theta \sqrt{1/12}, \varphi_i - \Delta \varphi \sqrt{1/12}), \quad \textbf{x}_{i}^{1}=(\theta_i - \Delta \theta \sqrt{1/12}, \varphi_i + \Delta \varphi \sqrt{1/12}),\\
&\textbf{x}_{i}^{2}=(\theta_i + \Delta \theta \sqrt{1/12}, \varphi_i - \Delta \varphi \sqrt{1/12}), \quad \textbf{x}_{i}^{3}=(\theta_i + \Delta \theta \sqrt{1/12}, \varphi_i + \Delta \varphi \sqrt{1/12}).
\end{align*}

{As a side note, designing a high-order numerical scheme with properties analogous to those presented in this paper within the framework of unstructured grids is not trivial. On the one hand, reconstruction operators, like CWENO,  become significantly more complex when applied to, for example, triangular meshes, as discussed in \cite{Dumbser2017}. On the other hand, working on this scenario would require a dedicated adaptation of the asymptotic reconstruction that will be introduced in the following Section~\ref{sec:4.2}, since the stencils associated with triangular cells differ substantially from those considered here.
}

\section{Geostrophic equilibrium and well-balanced reconstructions} \label{sec:4}

This section begins by deriving the expression of the geostrophic equilibrium in spherical coordinates through a linearization of system (\ref{2DRotatingSWESphere}) around a constant state where $h_{0}$, $u_{\theta}=0$, $u_{\varphi}=0$, assuming a flat bottom and constant atmospheric pressure. 
Next, we describe the steps to obtain well-balanced reconstruction operators that preserve the geostrophic equilibrium with high order of accuracy. 
A standard reconstruction operator is not inherently well-balanced, as the steady states that need to be preserved generally do not fit the polynomial representation that standard reconstruction operators can capture. However, following the strategies presented in \cite{Castro2008a, Castro2020}, it is possible to modify the standard operators to create well-balanced reconstruction operators.
Finally, we provide a proof of the asymptotic well-balanced behavior of the designed numerical schemes.


\subsection{Geostrophic equilibrium in spherical coordinates} \label{sec:4.1}
There are several methods for deriving the expression for the geostrophic equilibrium, both in cartesian and spherical coordinates. 
One approach is based on scale analysis: in the synoptic and mesoscale regimes, the non-linear advective terms in the rotating incompressible, inviscid, homogeneous Navier-Stokes system can be neglected \cite{Vallis2017, Zeitlin2018}, resulting in a stationary system where the Coriolis force is in balance with the pressure gradient, as shown in equation (\ref{cartesian_geostrophic}). 
Another approach is to linearize the 2D rotating NLSWE around a background state of constant positive height and zero velocities. This leads to a simplified PDE system, whose stationary states verify the geostrophic equilibrium \cite{Majda2012}. 
In this work, we will use this second approach to derive the expression for the geostrophic equilibrium in spherical coordinates.

Firstly, let's assume constant atmospheric pressure, $\nabla p^{a}=0$. 
We proceed by linearizing the system (\ref{2DRotatingSWESphere}) around a background state where $h_{0}=\mbox{cons}t > 0$, $u_{\theta}=0$, $u_{\varphi}=0$. 
Specifically, the unknown variables are expressed as $h=h_{0}+\epsilon \, \widetilde{h}$, $u_{\theta}=0+\epsilon \, \widetilde{u_{\theta}}$, $u_{\varphi}=0+\epsilon \, \widetilde{u_{\varphi}}$, where $\epsilon \ll 1$, indicating that the variables $\widetilde{h}$, $\widetilde{u_{\theta}}$, $\widetilde{u_{\varphi}}$ are small perturbations from the background state. 
We also assume small perturbations of a flat bottom $H= H_0 + \epsilon \, \widetilde{H}$. 
Next, in order to retrieve the conserved variables of system (\ref{2DRotatingSWESphere}), we first multiply $h$ by $\sigma$ to recover $h_{\sigma}= h_{0,\sigma}+\epsilon \, \widetilde{h_{\sigma}}$ and then multiply the velocities by $h_{\sigma}$ to obtain the discharges $Q_{\theta}=\epsilon \, h_{0,\sigma} \widetilde{u_{\theta}}$, $Q_{\varphi}=\epsilon \,  h_{0,\sigma} \widetilde{u_{\varphi}}$.
\noindent where the $\mathcal{O}(\epsilon^2)$ terms have been neglected. We now substitute these expressions into the system (\ref{2DRotatingSWESphere}) and again neglect the $\mathcal{O}(\epsilon^2)$ terms to derive the new linear system:
\begin{equation} \label{2DLinearRtotatingSWESphere}
\begin{cases}
\partial_{t}(\epsilon \widetilde{h_{\sigma}})+\frac{h_{0}}{R}\big[\partial_{\theta}(\epsilon \widetilde{u_{\theta}})+\partial_{\varphi}(\sigma \epsilon \widetilde{u_{\varphi}})\big]=0, \\
h_{0,\sigma}\partial_{t}(\epsilon \widetilde{u_{\theta}}) +\frac{gh_{0,\sigma}}{R\sigma}\partial_{\theta}(\epsilon \widetilde{h}) = \frac{gh_{0,\sigma}}{R\sigma}\partial_{\theta}(\epsilon\widetilde{H}) + h_{0,\sigma}f\epsilon \widetilde{u_{\varphi}}, \\
h_{0,\sigma}\partial_{t}(\epsilon \widetilde{u_{\varphi}}) +\frac{gh_{0,\sigma}}{R}\partial_{\varphi}(\epsilon \widetilde{h}) = \frac{gh_{0,\sigma}}{R}\partial_{\varphi}(\epsilon \widetilde{H}) - h_{0,\sigma}f\epsilon \widetilde{u_{\theta}}. 
\end{cases}
\end{equation}




\noindent Finally, the stationary states of the linearized 2D rotating spherical SWE system (\ref{2DLinearRtotatingSWESphere}) must be solutions of the system:
\begin{equation} \label{spherical_geostrophic}
\begin{cases}
\partial_{\theta}u_{\theta}+\partial_{\varphi}(\sigma u_{\varphi})=0, \\
g\partial_{\theta} \eta = Rf\sigma u_{\varphi}, \\
g\partial_{\varphi} \eta = -Rfu_{\theta},
\end{cases}
\end{equation}
\noindent where $\eta = h-H$, $\sigma=\mbox{cos} \, \varphi$ and $f=2\,\Omega \, \mbox{sin} \, \varphi$. 

The steady system (\ref{spherical_geostrophic}) represents the geostrophic equilibrium in spherical coordinates. 
The first equation accounts for the null divergence condition in the framework of spherical geometry, while the remaining two equations describe the balance between the pressure gradient and the Coriolis force, taking into account the specific influence of the spherical geometry.

\noindent It is important to emphasize that (\ref{spherical_geostrophic}) is not a steady state of system (\ref{2DRotatingSWESphere}) when $\nabla p^{a}=0$. 
However, (\ref{spherical_geostrophic}) governs the overall behavior of large-scale geophysical flows when the Rossby and Froude numbers are small. 
Moreover, it can be shown that the steady states of (\ref{2DRotatingSWESphere}) are close to the geostrophic equilibrium in this regime \cite{Majda2012}, which justifies the focus on the exact or approximate preservation of the geostrophic equilibrium.

\subsection{Asymptotic well-balancing} \label{sec:4.2}

The numerical scheme (\ref{semidiscreteNumericalSchemeUsingQuadrature}) presented in Section \ref{sec:3} is designed to provide a high-order approximation of the geostrophic equilibrium (\ref{spherical_geostrophic}) whenever a solution is close to it.
However, we cannot expect exact preservation of the geostrophic equilibrium, as it is not a steady state of system (\ref{2DRotatingSWESphere}). 
To address this, we first introduce the following definition of a exactly well-balanced reconstruction operator, which is extracted from \cite{Castro2020} and extended to the two-dimensional case.

\begin{definition} 
Given a stationary solution $U$ of the system of balance laws $$\partial_t U+\partial_x F_{1}(U) + \partial_y F_{2}(U) = S_{1}(U) \partial_x H + S_{2}(U) \partial_y H,$$ the reconstruction operator is said to be exactly well-balanced for $U$ if $$P_i(x,y;\{\overline{U}_j\}_{j\in\mathcal{S}_i})=U(x,y), \quad \forall (x,y) \in V_{i}=[x_{i-\frac{1}{2}},x_{i+\frac{1}{2}}]\times[y_{i-\frac{1}{2}},y_{i+\frac{1}{2}}], \, \forall i,$$ where $$\overline{U}_j=\frac{1}{|V_i|} \int_{V_i}U(x,y) \, dx \, dy, \quad \forall j,$$
or $\overline{U}_j$ is an approximation of the exact cell average given by a quadrature formula.
\end{definition}


The general well-balanced reconstruction strategy is presented in the following subsection.

\subsubsection{Well-balanced reconstructions} \label{WB_reconstructions}
Following the procedure described in \cite{Castro2020}, a standard reconstruction operator can be modified to make it well-balanced for the intended steady states to be preserved. 
We use the notation $\textbf{x}=(\theta,\varphi)$ and $P_i(\textbf{x})=P_i(\textbf{x};\{\overline{U}_j\}_{j\in\mathcal{S}_i})$ to represent a generic standard reconstruction operator applied to the cell averages $\overline{U}_j$ belonging to a particular stencil $\mathcal{S}_i$ (in our case, either MUSCL or CWENO). 
The reconstruction procedure adopts the following strategy: given the cell averages $\{h_{\sigma,j} \}$, $\{Q_{\theta,j} \}$, $\{Q_{\varphi,j} \}$, $\{H_{\sigma,j} \}$, $\{p_{\sigma,j}^{a} \}$, $\{\overline{\sigma}_j \}$, $j\in \mathcal{S}_i$

\medskip

\begin{enumerate}
    \item \textbf{Application of the reconstruction operator}.
    The reconstruction operator is applied to $\{h_{\sigma,j}\}$ and $\{p_{\sigma,j}^a\}$ to obtain the reconstructions $p_{h_{\sigma,i}}(\mathbf{x})$ and $p_{p^a_{\sigma,i}}(\mathbf{x})$ for the variable $h_{\sigma}$ and the given function $p_{\sigma}^a$ at cell $V_i$.
    
    \item \textbf{Definition of local steady state and fluctuations}. 
    Denote by $\eta_{\sigma}^*$, $Q_{\theta}^*$, and $Q_{\varphi}^*$ a local steady state of system (\ref{2DRotatingSWESphere}) at volume $V_i$, satisfying:
    \[ 
    \frac{1}{|V_i|}\int_{V_i} \eta_{\sigma}^* \, d\mathbf{x} = \eta_{\sigma,i}, \quad 
    \frac{1}{|V_i|}\int_{V_i} Q_{\theta}^* \, d\mathbf{x} = Q_{\theta,i}, \quad 
    \frac{1}{|V_i|}\int_{V_i} Q_{\varphi}^* \, d\mathbf{x} = Q_{\varphi,i}.
    \]
    If an analytical solution is not available, approximate strategies as described in \cite{GomezBueno2021} can be used to obtain approximations of the steady states. For $j \in \mathcal{S}_i$, compute the fluctuations:
    \begin{equation} \label{WBFluctuations}
    \begin{split}
    f_j^{\, \,\eta_{\sigma}} &= \eta_{\sigma,j}-\frac{1}{|V_j|}\int_{V_j}\eta_{\sigma}^{*} \, d\textbf{x},\\
    f_j^{\,Q_{\theta}} &= Q_{\theta,j}-\frac{1}{|V_j|}\int_{V_j}Q_{\theta}^{*} \, d\textbf{x},\\
    f_j^{\,Q_{\varphi}} &= Q_{\varphi,j}-\frac{1}{|V_j|}\int_{V_j}Q_{\varphi}^{*} \, d\textbf{x},
    \end{split}
    \end{equation}
    where the integrals are approximated using the corresponding quadrature rule.

    \item \textbf{Fluctuation reconstruction}. 
    Apply the reconstruction operator to the fluctuation values (\ref{WBFluctuations}) to define the fluctuation polynomials at cell $V_i$:
    \begin{equation} \label{FluctuationPolynomials}
    \begin{split}
    &p_{f_{i}^{\,\eta_{\sigma}}}(\textbf{x})=P_i(\textbf{x};\{f_{j}^{\, \,\eta_{\sigma}}\})\\
    &p_{f_{i}^{\,Q_{\theta}}}(\textbf{x}) =P_i(\textbf{x};\{f_{j}^{\,Q_{\theta}}\}), \\
    &p_{f_{i}^{\,Q_{\varphi}}}(\textbf{x})=P_i(\textbf{x};\{f_{j}^{\,Q_{\varphi}}\}).
    \end{split}
    \end{equation}

    \item \textbf{Reconstruction of variables}.
    Define the reconstruction functions for $\eta_{\sigma}$, $Q_{\theta}$, and $Q_{\varphi}$ at cell $V_i$ as:
    \begin{equation} \label{WBReconstruction}
    \begin{split}
    &p_{\eta_{\sigma},i}(\textbf{x})=p_{f_{i}^{\,\eta_{\sigma}}}(\textbf{x})+\eta_{\sigma}^{*}(\textbf{x}),\\
    &p_{Q_{\theta},i}(\textbf{x})=p_{f_{i}^{\,Q_{\theta}}}(\textbf{x})+Q_{\theta}^{*}(\textbf{x}),\\
    &p_{Q_{\varphi},i}(\textbf{x})=p_{f_{i}^{\,Q_{\varphi}}}(\textbf{x})+Q_{\varphi}^{*}(\textbf{x}).\\
    \end{split}
    \end{equation}

    \item \textbf{Reconstruction of $\widetilde{\eta_{\sigma}}$ and the modified bottom $\widetilde{H_{\sigma}}$}.
    The reconstruction of $\widetilde{\eta_{\sigma}}$ at cell $V_i$ is defined as $p_{\widetilde{\eta_{\sigma}},i}(\mathbf{x}) = p_{\eta_{\sigma},i}(\mathbf{x}) + p_{p_{\sigma,i}^a}(\mathbf{x}),$ and a high-order representation of the modified bottom $\widetilde{H_{\sigma}}$ is given by $p_{\widetilde{H_{\sigma}},i}(\mathbf{x}) = p_{h_{\sigma},i}(\mathbf{x}) - p_{\widetilde{\eta_{\sigma}},i}(\mathbf{x})$.
%


\end{enumerate}

\medskip

The following remark elaborates on the statement made at the end of Section \ref{sec:intro}, clarifying why this work expands upon and extends the results presented in \cite{Castro2017}.

\begin{remark} 
In absence of the Coriolis force and atmospheric pressure gradient, the described strategy yields reconstruction operators that are well-balanced for the steady states representing water at rest 
\begin{equation} \label{waterAtRest}
    u_{\theta}=0, \quad u_{\varphi}=0, \quad \eta_{\sigma}=\eta_{0}\, \mbox{cos}\, \varphi, \quad \eta_{0}=const.,
\end{equation}
 as, in this case, the fluctuation polynomials are identically zero for any solution of the form (\ref{waterAtRest}).
\end{remark}

One possible approach is to apply the well-balanced reconstruction strategy directly to the actual steady states of the system (\ref{2DRotatingSWESphere}). 
However, this approach is too complex as the steady states of (\ref{2DRotatingSWESphere}) are difficult to compute. Instead, based on the fact that the geostrophic equilibrium largely governs the behavior in regimes with low Froude and Rossby numbers, the idea is to obtain a local solution of (\ref{spherical_geostrophic}) and use it as a substitute for the steady state in the second step of the strategy to compute the fluctuations in (\ref{WBFluctuations}) and in the fourth step to define the reconstructions of $\eta_{\sigma}$, $Q_{\theta}$, and $Q_{\varphi}$ in (\ref{WBReconstruction}).
However, finding a local geostrophic equilibrium requires identifying a null divergence non-zero velocity field, which is not trivial. Therefore, the proposed approach is to seek high-order local approximations of the geostrophic solutions and use them as replacements for the steady state in the general strategy.

To put this idea into practice, two strategies for obtaining local approximations of the geostrophic equilibrium (\ref{spherical_geostrophic}) are outlined in the upcoming subsections. These strategies will be employed in the second-order and third-order schemes, respectively. While their implementations involve some minor differences, both strategies share the same fundamental approach: preserving a local high-order approximation of the velocity field and one order higher of the free surface, whose gradient is in balance with the Coriolis force. The main distinction between the two procedures lies in how the local velocity field is defined.

\subsubsection{Finding approximated local geostrophic equilibria for the sec\-ond-\-order scheme} \label{sec:4.2.2}
As explained in the previous subsection, we are looking for a approximated solution ($u_{\theta}^{*}$, $u_{\varphi}^{*}$, $\eta^{*}$) of (\ref{spherical_geostrophic}). For the implementation of the second-order scheme the strategy is to find a second-order approximation of the velocity field, i.e., set
\begin{equation} \label{localVelocityFieldSecondOrder}
\begin{cases}
u_{\theta}^{*}=u_0+u_1(\theta-\theta_{i})+u_2(\varphi-\varphi_i), \\
\sigma u_{\varphi}^{*}=v_0+v_1(\theta-\theta_{i})+v_2(\varphi-\varphi_i), \\
\end{cases}
\end{equation}
\noindent where $(\theta_i,\varphi_i)$ is the center of the volume $V_i$, and (\ref{spherical_geostrophic}) is used to derive the free-water surface expression:
\begin{equation} \label{localEtaSecondOrder}
\eta^{*}=C-\frac{Rf}{g} \bigg[u_0(\varphi-\varphi_i)+\frac{u_2}{2}(\varphi-\varphi_i)^2-v_0(\theta-\theta_i)-\frac{v_1}{2}(\theta-\theta_i)^2+u_1(\theta-\theta_i)(\varphi-\varphi_i)\bigg],
\end{equation}
\noindent where $f=2\, \Omega \, \mbox{sin} \varphi_i$ and we have used $v_2 = -u_1$ from the null divergence condition. Defining $\eta_{\sigma}^{*}=\eta^{*}\cdot \sigma$, the integration constant $C$ is fixed by imposing that the average value of $\eta_{\sigma}^{*}$ at cell $V_i$ must coincide with the average provided by the numerical scheme $\eta_{\sigma,i}$, i.e., $\frac{1}{|V_i|}\int_{V_i} \eta_{\sigma}^{*} \, d\textbf{x} = \eta_{\sigma,i}$, which leads using the mid-point quadrature rule to $C=\frac{\eta_{\sigma,i}}{\mbox{\small cos}\, \varphi_i}$.
\noindent Now, to approximate the velocity field $u_{\theta}^{*}, u_{\varphi}^{*}$, the coefficients $u_0,u_1,u_2,v_0,v_1$ are calculated minimizing the distance (least squares) between the averages of the linear velocity field in each volume of the computational stencil and the averages provided by the numerical method. Figure \ref{stencils} (left) shows a schematic representation of the stencil $\mathcal{S}_i$ associated to volume $V_i$. 
The procedure can be seen as imposing the conditions $\frac{1}{|V_j|}\int_{V_j} u_{\theta}^{*} \, d\textbf{x} = u_{\theta,j}$ and $\frac{1}{|V_j|}\int_{V_j} \sigma u_{\varphi}^{*} \, d\textbf{x} = \sigma_ju_{\varphi,j}$, for $j \in \mathcal{S}_i$, and using $v_2=-u_1$ to obtain a system $AX=B$ that is solved by least squares. The exact solution is:
\begin{equation}\label{localSolutionVelocityField}
\begin{cases}
u_0 = u_{\theta,0}, \quad u_1=\frac{1}{2(\Delta \theta ^2 + \Delta \varphi ^2)}\left[\Delta \theta(u_{\theta,3}-u_{\theta,2})+\Delta \varphi (\sigma_4 u_{\varphi,4}-\sigma_1 u_{\varphi,1})\right], \\ 
u_2 = \frac{1}{2 \Delta \varphi}(u_{\theta,1}-u_{\theta,4}), \quad v_0 = \sigma_0 u_{\varphi,0}, \quad  v_1 = \frac{1}{2 \Delta \theta}(\sigma_3 u_{\varphi,3}-\sigma_2 u_{\varphi,2}), \quad v_2 = -u_1, \\
\end{cases}
\end{equation}
\noindent where the subscripts in the RHS refer to the local numeration of the neighbor volumes.






\begin{figure}
    \centering
    \includegraphics[height=0.3\textwidth]{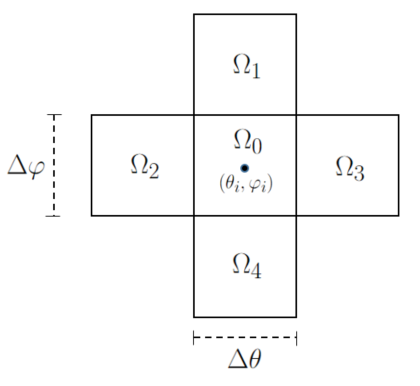}
    \hspace{2cm}
    \includegraphics[height=0.3\textwidth]{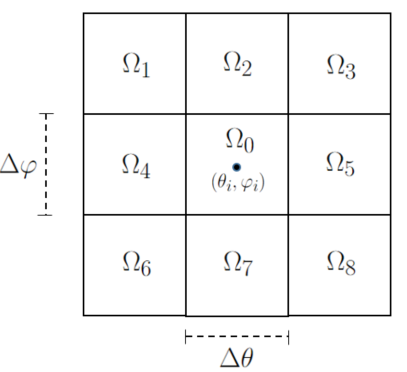}
    \caption{Stencil for the reconstruction operators. Second-order scheme (left) and third-order scheme (right). These are the local numeration, where $\Omega_0$ is the cell volume $V_i$. \label{stencils}}
\end{figure}

\subsubsection{Finding approximated local geostrophic equilibria for the third-order scheme} \label{sec:4.2.3}
For the implementation of the third-order scheme, the strategy to find an approximate solution ($u_{\theta}^{*}$, $u_{\varphi}^{*}$, $\eta^{*}$) of (\ref{spherical_geostrophic}) begins by obtaining a third-order approximation of the velocity field, i.e., we set:
\begin{equation} \label{localVelocityFieldThirdOrder}
\begin{cases}

u_{\theta}^{*}=u_0+u_1(\theta-\theta_{i})+u_2(\varphi-\varphi_i) \\
\quad \, \,  + \, \, u_3\big[(\theta-\theta_i)^2-\frac{\Delta \theta^2}{12}\big]+u_4\big[(\varphi-\varphi_i)^2-\frac{\Delta \varphi^2}{12}\big]+u_5(\theta-\theta_i)(\varphi-\varphi_i), \\

\sigma u_{\varphi}^{*}=v_0+v_1(\theta-\theta_{i})+v_2(\varphi-\varphi_i)\\
\quad \quad \, + \, v_3\big[(\theta-\theta_i)^2-\frac{\Delta \theta^2}{12}\big]+v_4\big[(\varphi-\varphi_i)^2-\frac{\Delta \varphi^2}{12}\big]+v_5(\theta-\theta_i)(\varphi-\varphi_i), \\
\end{cases}
\end{equation}

\noindent where $(\theta_i,\varphi_i)$ is the center of the volume $V_i$, and we use (\ref{spherical_geostrophic}) to derive the free-water surface expression:

\begin{equation} \label{localEtaThirdOrder}
\begin{split}
\eta^{*}  =   C - \frac{Rf}{g} & \bigg[\bigg(-v_0+v_3\frac{\Delta \theta^2}{12}-u_5\frac{\Delta \varphi^2}{24}\bigg)(\theta-\theta_i)\\
& + \bigg(u_0-u_3\frac{\Delta \theta^2}{12}-u_4\frac{\Delta \varphi^2}{12}\bigg)(\varphi-\varphi_i)+ \frac{u_2}{2}(\varphi-\varphi_i)^2\\
&-\frac{v_1}{2}(\theta-\theta_i)^2+u_1(\theta-\theta_i)(\varphi-\varphi_i)-\frac{v_3}{3}(\theta-\theta_i)^3\\
& +\frac{u_4}{3}(\varphi-\varphi_i)^3+u_3(\theta-\theta_i)^2(\varphi-\varphi_i)+\frac{u_5}{2}(\theta-\theta_i)(\varphi-\varphi_i)^2\bigg],
\end{split}
\end{equation}

\noindent where $f=2\, \Omega \, \mbox{sin} \varphi_i$ and the relations 
\begin{equation} \label{DN_conditionsThirdOrder}
v_2=-u_1,\quad v_4=-\frac{u_5}{2},\quad v_5=-2u_3,
\end{equation}

 \noindent obtained from the null divergence condition have been used. 
 Note that for the definition of (\ref{localVelocityFieldThirdOrder}) we have used a basis of the space of second-degree polynomials $\mathbb{P}_2(\theta,\varphi)$ which satisfies the condition that all non-constant basis elements have zero mean in the central cell $\Omega_0$ \cite{CastroDiaz2018}. 
 Defining again $\eta_{\sigma}^{*}=\eta^{*}\cdot \sigma$, the integration constant $C$ is determined by imposing that the average value of $\eta_{\sigma}^{*}$ at cell $V_i$ must coincide with the average provided by the numerical scheme $\eta_{\sigma,i}$. 
 Using the two-point Gauss quadrature formula, we obtain the value: 
\begin{align*}
C=\frac{\eta_{\sigma,i}}{\overline{\sigma}_i}+\frac{Rf}{216\,g\,\overline{\sigma}_i} & \bigg[\frac{9}{2}\big(\Delta \varphi^2u_2-\Delta \theta^2v_1\big)\bigg(\mbox{\small cos}\big(\varphi_i+\frac{\Delta \varphi}{\sqrt{12}}\big)+\mbox{\small cos}\big(\varphi_i-\frac{\Delta \varphi}{\sqrt{12}}\big)\bigg)+\\
& +\sqrt{3}\Delta \varphi(-18u_0+\Delta\varphi^2u_4)\bigg(\mbox{\small cos}\big(\varphi_i+\frac{\Delta \varphi}{\sqrt{12}}\big)-\mbox{\small cos}\big(\varphi_i-\frac{\Delta \varphi}{\sqrt{12}}\big)\bigg)\bigg].
\end{align*}

\noindent Figure \ref{stencils} (right) shows a schematic representation of the stencil $\mathcal{S}_i$ associated with volume $V_i$. The coefficients for the local velocity field are computed differently than in the second-order case. 
We have attempted to apply the same procedure to the third-order  scheme as we did in the second-order case, but it led to significant instabilities when starting close to a geostrophic equilibrium. 
This issue likely stems from the poor conditioning of the matrix $A^{T}A$ when solving the system $AX=B$ by least squares, analogously to the second-order case. 
In the second-order case, the matrix $A^{T}A$ has a constant condition number $\mbox{cond}_{2}(A^{T}A)=2$ when $\Delta\theta=\Delta\varphi$, meaning that it does not significantly propagate errors in the approximation of the velocity field. 
However, this is not true for the third-order case, where $\mbox{cond}_{2}(A^{T}A)$ increases by a factor of 4 when halving the spatial resolution.
{To address this issue, we choose to apply a CWENO reconstruction individually to each velocity component, $u_{\theta}^{*}$ and $u_{\varphi}^{*}$, and then enforce the null-divergence condition by imposing the relations (\ref{DN_conditionsThirdOrder}) on the coefficients of the resulting polynomials. Specifically, we first apply the CWENO reconstruction procedure to the cell averages $\{u_{\theta,j}\}_{j \in \mathcal{S}_i}$ and $\{\sigma_j u_{\varphi,j}\}_{j \in \mathcal{S}_i}$ to obtain quadratic reconstruction polynomials $p^{\prime}_{u_{\theta},i}$, $p^{\prime}_{\sigma u_{\varphi},i}$, resp., of the form:
{\small
\begin{align*}
p^{\prime}_{u_{\theta},i}(\theta,\varphi) &= u^{\prime}_0 + u^{\prime}_1(\theta - \theta_i) + u^{\prime}_2(\varphi - \varphi_i) 
+ u^{\prime}_3\left[(\theta - \theta_i)^2 - \frac{\Delta \theta^2}{12}\right] \\
&\quad + u^{\prime}_4\left[(\varphi - \varphi_i)^2 - \frac{\Delta \varphi^2}{12}\right] 
+ u^{\prime}_5(\theta - \theta_i)(\varphi - \varphi_i),
\end{align*}
}
\vspace{-1em}
%
{\small
\begin{align*}
p^{\prime}_{\sigma u_{\varphi},i}(\theta,\varphi) &= v^{\prime}_0 + v^{\prime}_1(\theta - \theta_i) + v^{\prime}_2(\varphi - \varphi_i) 
+ v^{\prime}_3\left[(\theta - \theta_i)^2 - \frac{\Delta \theta^2}{12}\right] \\
&\quad + v^{\prime}_4\left[(\varphi - \varphi_i)^2 - \frac{\Delta \varphi^2}{12}\right] 
+ v^{\prime}_5(\theta - \theta_i)(\varphi - \varphi_i).
\end{align*}
}
\par\noindent 
Finally, we define $u_{\theta}^{*}$ and $\sigma u_{\varphi}^{*}$ as in (\ref{localVelocityFieldThirdOrder}), assigning the coefficients not involved in the conditions (\ref{DN_conditionsThirdOrder}) as follows: $u_0 = u^{\prime}_0$, $u_2 = u^{\prime}_2$, $u_4 = u^{\prime}_4$, $v_0 = v^{\prime}_0$, $v_1 = v^{\prime}_1$, and $v_3 = v^{\prime}_3$. The remaining coefficients are determined using an approximation criterion that guarantees the satisfaction of relations (\ref{DN_conditionsThirdOrder}). The use of the CWENO reconstruction here explains why we did not use the canonical polynomial basis to construct the local velocity field in~(\ref{localVelocityFieldThirdOrder}).
}
\subsubsection{Asymptotic well-balanced behavior}\label{sec:4.2.4}

Following the strategy outlined above, which combines the general reconstruction procedure with the computed local geostrophic equilibrium as a substitute for a local stationary solution, we can derive the following asymptotic behavior property of the second-order numerical scheme.

\begin{theorem} \label{theo:AWB}
In the case of a flat bottom $\nabla H=0$ and constant in space atmospheric pressure $\nabla p^{a}=0$, the numerical scheme (\ref{semidiscreteNumericalSchemeUsingQuadrature}) along with the reconstruction functions defined in subsection \ref{WB_reconstructions} is asymptotically well-balanced for the geostrophic equilibrium (\ref{spherical_geostrophic}) in the following sense: Given the initial condition $w(\textbf{x},0)=w_{0}^{*}(\textbf{x})=(h_{\sigma}^{*}(\textbf{x}),h_{\sigma}^{*}(\textbf{x})u_{\theta}^{*}(\textbf{x}), h_{\sigma}^{*}(\textbf{x})u_{\varphi}^{*}(\textbf{x}))$, where  $h_{\sigma}^{*}(\textbf{x}) = h^{*}(\textbf{x}) \cdot \sigma(\varphi) $ and 
\begin{equation} \label{localGeostrophicOrderEps}
\begin{cases}
    \, \, \, \, u_{\theta}^{*} =\epsilon\big[ u_0+u_1(\theta-\theta_i)+u_2(\varphi-\varphi_i)\big], \\
    \sigma u_{\varphi}^{*}=\epsilon\big[ v_0+v_1(\theta-\theta_i)-u_1(\varphi-\varphi_i)\big], \\
    \, \, \, \, \, h^{*}=h_i+\epsilon \frac{Rf}{g}\big[v_0(\theta-\theta_i)+\frac{v_1}{2}(\theta-\theta_i)^2-u_0(\varphi-\varphi_i)\\
    \quad \quad \quad \quad \quad \quad \quad \, -\frac{u_2}{2}(\varphi-\varphi_i)^2-u_1(\theta-\theta_i)(\varphi-\varphi_i)\big].
\end{cases}
\end{equation}

\noindent then $w_i^{'}(0)=\mathcal{O}(\epsilon^2) \, \, \forall i$, that is, the approximated geostrophic equilibria (\ref{localVelocityFieldSecondOrder})-(\ref{localEtaSecondOrder})  are also approximated discrete equilibria.
\end{theorem}

{

\noindent \textsc{Proof}: Based on the hypothesis of the theorem, system (\ref{2DRotatingSWESphereCompactVectorSolvedSystem}) can be rewritten as 
\begin{equation}
\partial_{t}w+\frac{1}{R}\big[\partial_{\theta}F_{\theta}(W)+\partial_{\varphi}F_{\varphi}(W)+G_{\varphi}^{1}(W)\partial_{\varphi}\sigma-G_{\varphi}^{2}(W)\partial_{\varphi}\sigma\big]=0,
\end{equation}

\noindent being

$$W=\left( \begin{array}{c} w \\ \sigma \end{array} \right), \quad w=\left( \begin{array}{c} h_{\sigma} \\ Q_{\theta} \\ Q_{\varphi} \end{array} \right), \quad {F}_{\theta}(W)=\left( \begin{array}{c} \frac{Q_{\theta}}{\sigma} \\ \frac{Q_{\theta}^2}{h_{\sigma}\sigma}+\frac{gh_{\sigma}^2}{2\sigma^2} \\ \frac{Q_{\theta}Q_{\varphi}}{h_{\sigma}\sigma} \end{array} \right),$$

$${F}_{\varphi}(W)=\left( \begin{array}{c} Q_{\varphi} \\ \frac{Q_{\theta}Q_{\varphi}}{h_{\sigma}} \\ \frac{Q_{\varphi}^2}{h_{\sigma}}+\frac{gh_{\sigma}^2}{2\sigma}  \end{array} \right), \, \, G_{\varphi}^{1}(W)=\left( \begin{array}{c} 0 \\ \frac{Q_{\theta}Q_{\varphi}}{h_{\sigma}\sigma}+2\Omega Q_{\varphi}R \\ -\frac{Q_{\theta}^2}{h_{\sigma}\sigma}-2\Omega Q_{\theta}R \end{array} \right), \, \, G_{\varphi}^{2}(W)=\left( \begin{array}{c} 0 \\ 0 \\ \frac{gh_{\sigma}^2}{2\sigma^2} \end{array} \right).$$
\noindent where it has been used that the non-zero component $\frac{gh_{\sigma}}{\sigma}\partial_{\varphi}h_{\sigma}$ in the pressure term $T_{\varphi}^{p}(W) \partial_{\varphi}h_{\sigma}$ can be written as $$\frac{gh_{\sigma}}{\sigma}\partial_{\varphi}h_{\sigma}=\frac{g}{2}\partial_{\varphi}\bigg(\frac{(h_{\sigma})^2}{\sigma} \bigg)+\frac{gh^2}{2}\partial_{\varphi}\sigma.$$ 
Thus, the first term is added to the flux in latitude $F_{\varphi}(W)$ and the second term is incorporated into the third component of $G_{\varphi}^2(h_{\sigma},h_{\sigma},\sigma)\partial_{\varphi}\sigma$. (Note that we are not working with the dimensionless system in order to maintain more generality). 

The semidiscrete in space numerical scheme (\ref{semidiscreteNumericalScheme}) is expressed as:
\begin{equation} \label{simplifiedNumericalScheme}
    w_i'(t)=\frac{-1}{R|V_i|}\bigg[ \sum_{j\in\mathcal{N}_i} \int_{E_{i,j}} \mathbb{F}(W_{i,j}^{-}(\gamma),W_{i,j}^{+}(\gamma),\vec{n}_{i,j}) d\gamma +S_i\bigg]
\end{equation}
\noindent where $\mathbb{F}(W_L,W_R,\vec{n_{i,j}})=F_{\vec{n_{i,j}}}(W_R)-\delta_{i,j}(\sigma) \mathcal{D}^{-}_{i,j}(W_L,W_R)$ is a numerical flux consistent with $F_{\vec{n_{i,j}}}$ and $S_i$ is an approximation of the integral of the source term $S_i \approx \int_{V_i} \big[G_{\varphi}^{1}(P_i(\textbf{x}))-G_{\varphi}^{2}(P_i(\textbf{x})) \big]\partial_{\varphi}\sigma \, d\textbf{x}$.
Note that we are assuming a constant bottom, so there is no contribution from variations in the topography to the line integrals at the edges of the cell.

Let $w_{0}^{*}(\textbf{x})$ be the initial condition in the hypothesis of the theorem; i.e.,  it satifies equation (\ref{localGeostrophicOrderEps}). 
Define $W_i^{*}(\textbf{x})=W_0(\textbf{x})=(w_{0}^{*}(\textbf{x}),\sigma(\textbf{x}))$ as the restriction of the initial condition to the cell $V_i$. 
Denote $Q_{\theta}^{*}(\textbf{x})=h_{\sigma}^{*}(\textbf{x})u_{\theta}^{*}(\textbf{x})$ and 
$Q_{\varphi}^{*}(\textbf{x})=h_{\sigma}^{*}(\textbf{x})u_{\varphi}^{*}(\textbf{x})$. 
From equation (\ref{localGeostrophicOrderEps}), the following identities hold:

\begin{equation} 
\begin{cases}
    h_{\sigma}^{*}=h^{*}\sigma = h_{i,\sigma}+\mathcal{O}(\epsilon), \\
    Q_{\theta}^{*}=h_{\sigma}^{*}u_{\theta}^{*}=h_{i,\sigma}u_{\theta}^{*} + \mathcal{O}(\epsilon^2), \\
    Q_{\varphi}^{*}=h_{\sigma}^{*}u_{\varphi}^{*}=h_i(\sigma u_{\varphi}^{*}) + \mathcal{O}(\epsilon^2).
\end{cases}
\end{equation}
\noindent where $h_{i,\sigma}:=h_{i}\,\sigma$.

\noindent From the geostrophic relations (\ref{spherical_geostrophic}), it is straightforward to check that: 
\begin{equation} \label{neccesaryEqualities}
\begin{cases}
    u_{\theta}^{*}=-\frac{g}{fR}\partial_{\varphi}h^{*}, \\
    u_{\varphi}^{*}=\frac{g}{fR\sigma}\partial_{\theta}h^{*}, \\
    g h_{\sigma}^{*}\partial_{\varphi}h^{*}=\frac{g}{2}\bigg[\partial_{\varphi}\bigg(\frac{(h_{\sigma}^{*})^2}{\sigma}\bigg)-(h^{*})^2\partial_{\varphi}\sigma \bigg]
\end{cases}
\end{equation}

Recalling the definition of the reconstruction operator $P_i(\textbf{x})$ from subsection \ref{WB_reconstructions}, we have $P_i(\textbf{x})=W_i^{*}(\textbf{x})$, since the fluctuation polynomials (\ref{FluctuationPolynomials}) are initially all equal to zero. 
Operating with the source term, we obtain:
\begin{equation} \label{sourceTermsCase1}
\begin{split}
        & \frac{1}{R}\left[ G_{\varphi}^{1}(P_i(\textbf{x}))\partial_{\varphi}\sigma-G_{\varphi}^{2}(P_i(\textbf{x})) \partial_{\varphi}\sigma\right] = \frac{1}{R}\left[G_{\varphi}^{1}(W_{i}^{*}(\textbf{x}))\partial_{\varphi}\sigma-G_{\varphi}^{2}(W_{i}^{*}(\textbf{x})) \partial_{\varphi}\sigma)\right]  \\
        = &\frac{1}{R}\left[ \left( \begin{array}{c} 0 \\ 2\,\Omega \, Q_{\varphi}^{*} R \partial_{\varphi}\sigma + \mathcal{O}(\epsilon^2) \\ -2\,\Omega \, Q_{\theta}^{*} R \partial_{\varphi}\sigma + \mathcal{O}(\epsilon^2)\end{array} \right) - \left( \begin{array}{c} 0 \\ 0 \\ \frac{g(h_{\sigma}^{*} )^2}{2\sigma^2}\partial_{\varphi}\sigma\end{array} \right) \right] \\
        = &\frac{1}{R}\left[ \left( \begin{array}{c} 0 \\ -fRQ_{\varphi}^{*} + \mathcal{O}(\epsilon^2) \\ fRQ_{\theta}^{*} + \mathcal{O}(\epsilon^2)\end{array} \right) - \left( \begin{array}{c} 0 \\ 0 \\ \frac{g(h_{\sigma}^{*} )^2}{2\sigma^2}\partial_{\varphi}\sigma\end{array} \right)\right]  \\
        = &\frac{1}{R}\left[ \left( \begin{array}{c} 0 \\ -\frac{g}{2}\partial_{\theta}\big( (h^{*})^2\big) + \mathcal{O}(\epsilon^2) \\ -\frac{g}{2}\partial_{\varphi}\bigg( \frac{(h_{\sigma}^{*})^2}{\sigma}\bigg)+\frac{g}{2}(h^{*})^2\partial_{\varphi}\sigma + \mathcal{O}(\epsilon^2)\end{array} \right) - \left( \begin{array}{c} 0 \\ 0 \\ \frac{g(h_{\sigma}^{*} )^2}{2\sigma^2}\partial_{\varphi}\sigma\end{array} \right) \right]  \\
        = & \frac{1}{R}\left( \begin{array}{c} 0 \\ -\frac{g}{2}\partial_{\theta}\big( (h^{*})^2\big) + \mathcal{O}(\epsilon^2) \\ -\frac{g}{2}\partial_{\varphi}\bigg( \frac{(h_{\sigma}^{*})^2}{\sigma}\bigg)+ \mathcal{O}(\epsilon^2)\end{array} \right) = \frac{1}{R} \left( \begin{array}{c} 0 \\ -\frac{g}{2\sigma}\partial_{\theta}\big( \frac{(h_{\sigma}^{*})^2}{\sigma}\big) + \mathcal{O}(\epsilon^2) \\ -\frac{g}{2}\partial_{\varphi}\bigg( \frac{(h_{\sigma}^{*})^2}{\sigma}\bigg)+ \mathcal{O}(\epsilon^2)\end{array} \right) \\
        = & \left( \begin{array}{c} 0 \\ -\frac{g}{2}\nabla_s\Psi(h_{\sigma}^{*},\sigma)+\mathcal{O}(\epsilon^2) \end{array} \right),
\end{split}
\end{equation}

\noindent where $\Psi(h_{\sigma}^{*},\sigma)= \frac{(h_{\sigma}^{*})^2}{\sigma}$; $\nabla_s=\frac{1}{R}(\frac{1}{\sigma}\frac{\partial}{\partial \theta},\frac{\partial}{\partial \varphi})$ is the surface gradient in spherical coordinates and the identities (\ref{neccesaryEqualities}) have been used. Expression (\ref{sourceTermsCase1}) allows us to rewrite the integral of the source term as:
\begin{equation}
\begin{split}
        \frac{1}{R} \int_{V_i} \big[G_{\varphi}^{1}(W_{i}^{*}(\textbf{x}))-G_{\varphi}^{2}(W_{i}^{*}(\textbf{x}))  \big] & \partial_{\varphi}\sigma \, d\textbf{x} = \left( \begin{array}{c} 0 \\ \displaystyle \int_{V_i} -\frac{g}{2}\nabla_s\Psi(h_{\sigma}^{*},\sigma)\, d\textbf{x}+\mathcal{O}(\epsilon^2) \end{array} \right) = \\
        = & \left( \begin{array}{c} 0 \\ \displaystyle \sum_{j\in \mathcal{N}_i} \int_{E_{i,j}} -\frac{g}{2}\frac{h_{\sigma}^{*}(\gamma)^2}{\sigma(\gamma)} \frac{n^{\theta}_{i,j}}{R \sigma(\gamma)} \, d\gamma + \mathcal{O}(\epsilon^2) \\ \displaystyle \sum_{j\in \mathcal{N}_i} \int_{E_{i,j}}  -\frac{g}{2}\frac{h_{\sigma}^{*}(\gamma)^2}{\sigma(\gamma)} \frac{n^{\varphi}_{i,j}}{R} \, d\gamma + \mathcal{O}(\epsilon^2)\end{array} \right).
\end{split}
\end{equation}

We can approximate this integral of the source terms by applying the mid-point quadrature formula, which gives the approximation:
\begin{equation} \label{approximationSourceTerms}
        \frac{1}{R} \, \Tilde{S}_i = \left( \begin{array}{c} 0 \\ \displaystyle \sum_{j\in \mathcal{N}_i} |E_{i,j}| (-\frac{g}{2})\frac{h_{\sigma}^{*}(\textbf{x}_{ij})^2}{\sigma(\textbf{x}_{ij})} \frac{n^{\theta}_{i,j}(\textbf{x}_{ij})}{R\sigma(\textbf{x}_{ij})}  + \mathcal{O}(\epsilon^2) \\ \displaystyle \sum_{j\in \mathcal{N}_i} |E_{i,j}|  (-\frac{g}{2})\frac{h_{\sigma}^{*}(\textbf{x}_{ij})^2}{\sigma(\textbf{x}_{ij})} \frac{n^{\varphi}_{i,j}(\textbf{x}_{ij})}{R}  + \mathcal{O}(\epsilon^2)\end{array} \right).
\end{equation}

\noindent where $\textbf{x}_{ij}$ is the center of the edge $E_{i,j}$.

We have derived an expression for the approximation $\Tilde{S}_i$ of the source terms $S_i$ in (\ref{approximationSourceTerms}). The next final step is to approximate the line integrals of the numerical flux in equation (\ref{simplifiedNumericalScheme}). 
Since we already know that $P_i(\textbf{x})=W_i^{*}(\textbf{x})$, it follows that $W_{i,j}^{-}=W_{i,j}^{+}=W_{0}(\gamma)$ due to continuity. 
Therefore, the consistency of the numerical flux implies that $\mathbb{F}(W_{i,j}^{-},W_{i,j}^{+},\vec{n}_{i,j})=\mathbb{F}(W_{0}(\gamma),W_{0}(\gamma),\vec{n}_{i,j})=F_{\vec{n}}(W_{0}(\gamma))=  \left( \begin{array}{c} F_{\vec{n}}^{1}(W_{0}(\gamma)) \\ F_{\vec{n}}^{2}(W_{0}(\gamma)) \end{array} \right)$.

On one hand, $F_{\vec{n}}^{1}(W)=n_{\theta}\big( \frac{Q_{\theta}}{\sigma}\big) + n_{\varphi}Q_{\varphi}$, so we have $F_{\vec{n}}^{1}(W_{0}(\gamma))=h_i\big[ n_{\theta}u_{\theta}^{*}+n_{\varphi}\sigma u_{\varphi}^{*}\big]+\mathcal{O}(\epsilon^2)$. Thus, we have 
\begin{equation}
\begin{split}
\frac{1}{R} \sum_{j \in \mathcal{N}_i} \int_{E_{i,j}} F_{\vec{n}}^{1}(W_{0}(\gamma)) \, d\gamma & =  \frac{h_i}{R} \sum_{j \in \mathcal{N}_i} \int_{E_{i,j}} (n_{\theta}u_{\theta}^{*}+n_{\varphi}\sigma u_{\varphi}^{*}) \, d\gamma +\mathcal{O}(\epsilon^2) \\
&=\frac{h_i}{R} \int_{V_i} \nabla \cdot (u_{\theta}^{*},\sigma u_{\varphi}^{*}) \, d\textbf{x} +\mathcal{O}(\epsilon^2) = \mathcal{O}(\epsilon^2).
\end{split}
\end{equation}

On the other hand, $$F_{\vec{n}}^{2}(W) = \left( \begin{array}{c} n_{\theta}\big( \frac{Q_{\theta}^2}{h_{\sigma}\sigma}\big) + n_{\theta}\frac{gh_{\sigma}^2}{2\sigma^2}+n_{\varphi}\frac{Q_{\theta}Q_{\varphi}}{h_{\sigma}} \\ n_{\theta}\frac{Q_{\theta}Q_{\varphi}}{h_{\sigma}\sigma}+n_{\varphi}\frac{Q_{\varphi}^2}{h_{\sigma}}+n_{\varphi}\frac{gh_{\sigma}^2}{2\sigma} \end{array} \right),$$ 
so $$F_{\vec{n}}^{2}(W_{0}(\gamma)) = \left( \begin{array}{c} n_{\theta}\frac{g(h_{\sigma}^{*})^2}{2\sigma^2}+\mathcal{O}(\epsilon^2) \\ n_{\varphi}\frac{g(h_{\sigma}^{*})^2}{2\sigma}+\mathcal{O}(\epsilon^2) \end{array} \right).$$

\noindent Finally, using again the mid-point quadrature formula, we obtain the desired result:

\begin{equation} \notag
    \begin{split}
         w_i'(0)&=\frac{-1}{R|V_i|}\bigg[ \sum_{j\in\mathcal{N}_i} |E_{i,j}| \mathbb{F}(W_{i,j}^{-}(\textbf{x}_{ij}),W_{i,j}^{+}(\textbf{x}_{ij}),\vec{n}_{i,j}) +\Tilde{S}_i\bigg] \\
        & = \frac{-1}{|V_i|}  \left[ \frac{1}{R}\begin{array}{c}  \left( \begin{array}{c}\displaystyle \sum_{j \in \mathcal{N}_i} |E_{i,j}| F_{\vec{n}}^{1}(W_{0}(\textbf{x}_{ij})) \\ \displaystyle\sum_{j \in \mathcal{N}_i} |E_{i,j}| F_{\vec{n}}^{2}(W_{0}(\textbf{x}_{ij}))  \end{array} \right) + \frac{1}{R} \, \Tilde{S}_i \end{array} \right] \\
        & =  \frac{-1}{|V_i|} \left[ \begin{array}{c} \left( \begin{array}{c} \mathcal{O}(\epsilon^2) \\ \displaystyle \sum_{j \in \mathcal{N}_i} |E_{i,j}| \frac{g}{2}\frac{h_{\sigma}^{*}(\textbf{x}_{ij})^2}{\sigma(\textbf{x}_{ij})}\frac{n^{\theta}_{i,j}(\textbf{x}_{ij})}{R \sigma(\textbf{x}_{ij})} + \mathcal{O}(\epsilon^2) \\ \displaystyle \sum_{j \in \mathcal{N}_i} |E_{i,j}| \frac{g}{2}\frac{h_{\sigma}^{*}(\textbf{x}_{ij})^2}{\sigma(\textbf{x}_{ij})}\frac{n^{\varphi}_{i,j}(\textbf{x}_{ij})}{R} + \mathcal{O}(\epsilon^2) \end{array} \right) + \frac{1}{R} \, \Tilde{S}_i \end{array} \right] = \mathcal{O}(\epsilon^2),
    \end{split}
\end{equation}

\noindent where the expression for the source term (\ref{approximationSourceTerms}) has been used $\Box$


}

\begin{remark} 
The analogous version of \textbf{Theorem 1} for the case of a third-order approximation of the geostrophic equilibrium is also true.
Specifically, under the hypothesis of \textbf{Theorem 1}, the approximated geostrophic equilibria given by (\ref{localVelocityFieldThirdOrder})-(\ref{localEtaThirdOrder}) are also approximated discrete equilibria. 
The proof is similar, but it is not included for the sake of simplicity.
\end{remark}

\section{Numerical results} \label{sec:5}


In this final section, we present the numerical results obtained from the designed numerical schemes. Four different schemes are considered for comparison. Two of them are non-asymptotically well-balanced, as the reconstructed coefficients of the local geostrophic equilibrium in (\ref{localVelocityFieldSecondOrder}) or (\ref{localVelocityFieldThirdOrder}) are set to zero. 
These schemes are labeled Geos-NAWB2 and Geos-NAWB3 for the second- and third-order cases, respectively. 
The other two schemes incorporate the complete reconstruction procedure for the local geostrophic equilibrium and are referred to as Geos-AWB2 and Geos-AWB3 for the second- and third-order cases, respectively. 
The CFL condition is given by: $$ \Delta t = CFL \,  \min_{i} \bigg(\frac{R \Delta \theta \Delta \varphi \, \mbox{cos}\, \varphi_i}{(|u_{\theta,i}|+\sqrt{gh_i})\Delta \varphi + (|u_{\varphi,i}|+\sqrt{gh_i})\Delta \theta}\bigg),$$ 
where $\Delta \theta$ and $\Delta \varphi$ are the mesh sizes in the spatial directions. A parallel GPU implementation is performed to accelerate the computations and reduce the overall computational time. 


\subsection{Stationary vortex on Cartesian coordinates}
This test is taken from \cite{Chertock2017}. 
Similar tests can be found in \cite{Audusse2011, Audusse2018}. 
To assess the ability of the numerical schemes to preserve the geostrophic equilibrium structure, we first reduce the problem to a Cartesian framework, taking as initial condition a stationary solution of the 2D rotating NLSWE. 
Specifically, we consider a stationary vortex in the squared domain $[-1,1]\times[-1,1]$ with the boundary conditions set to be a zero-order extrapolation in both $x$- and $y$-directions. 
We assume a flat atmospheric pressure $\nabla p^{a} = 0$, flat bottom topography $H = 0$, and constant Coriolis parameter $f = \frac{1}{\epsilon}$.
The gravity acceleration is set to $g = \frac{1}{\epsilon^2}$, with $\epsilon = 0.05$. 
Note that we are in the regime of small and of the same order for the Froude and Rossby numbers. i.e., $Fr \approx Ro \approx \epsilon=5 \cdot 10^{-2}$.
The initial conditions are given by:
\begin{equation} \notag
    \begin{split}
         & h(x,y,0)=1+\epsilon^2 \Psi(\sqrt{x^2+y^{\,2}}), \\
         & u(x,y,0)=-\epsilon y \, \xi(\sqrt{x^2+y^{\,2}}), \\
         & v(x,y,0)=\epsilon x \, \xi(\sqrt{x^2+y^{\,2}}),
    \end{split}
\end{equation}

\noindent where

\begin{small}
\begin{equation}
    \Psi(r) =
    \begin{cases}
        \begin{aligned}
            &\quad \quad \frac{5}{2}(1+5\epsilon^2)r^{\,2}, & &r < \frac{1}{5} \\
            \frac{1+5\epsilon^2}{10} + 2r - \frac{3}{10} & - \frac{5}{2}r^{\,2} + \epsilon^2\bigg(4 \log(5r) + \frac{7}{2} - 20r + \frac{25}{2}r^{\,2}\bigg), & &\frac{1}{5} \leq r < \frac{2}{5} \\
            &\frac{1 - 10\epsilon^2 + 20\epsilon^2 \log(2)}{5}, & &r \geq \frac{2}{5}
        \end{aligned}
    \end{cases}
\end{equation}
\end{small}

\begin{small}
\begin{equation}
    \xi(r) =
    \begin{cases}
        \begin{aligned}
            &5, & &r < \frac{1}{5} \\
            \frac{2}{r} &- 5, & &\frac{1}{5} \leq r < \frac{2}{5} \\
            &0, & &r \geq \frac{2}{5}
        \end{aligned}
    \end{cases}
\end{equation}
\end{small}

\begin{figure}[h]
    \centering
    \includegraphics[height=0.5\textwidth]{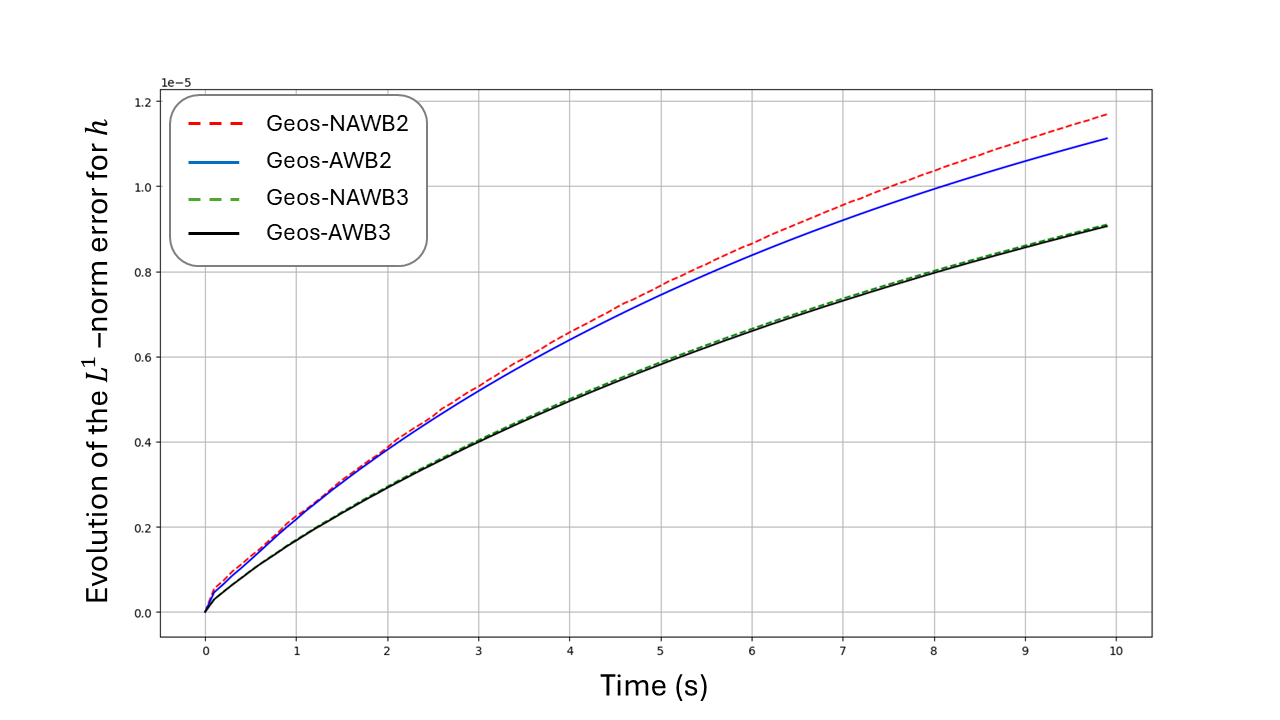}

    \caption{Evolution of water height errors in $L^{1}$-norm for the stationary vortex test case. Dashed lines represent the non-asymptotically well-balanced schemes, while solid lines correspond to the asymptotically well-balanced schemes, where the complete local geostrophic reconstruction is performed. \label{errors_test_chertock}}.
\end{figure}

We compute the numerical solutions using both the asymptotically well-balanced (Geos-AWB2, Geos-AWB3) and non-asymptotically well-balanced (Geos-NAWB2, Geos-NAWB3) schemes. Exact preservation of the equilibrium cannot be expected for two reasons.

First, this solution is not a geostrophic equilibrium; it is a stationary solution of the 2D rotating NLSWE that converges to the geostrophic equilibrium as $\epsilon$ tends to zero. 
Although, as $\epsilon$ is already small, we are very close to the geostrophic equilibrium. 
Second, the second-order (or third-order) scheme preserves a second-order (or third-order) approximation of the velocity field in a geostrophic equilibrium. 
So again, no exact preservation should be expected. 
However, the goal of the reconstruction procedure is to maintain the structure of this stationary vortex as long as possible. 

The domain is divided into an uniform structured grid of $100\times100$ volumes. 
The simulation is run for $T=10$s, with results saved each $0.1$s. 
Since the exact solution is known, the evolution of the error in the water height $h$ relative to the exact solution is computed in $L^{1}$-norm for the four numerical schemes. 
The results, shown in Figure \ref{errors_test_chertock} demonstrate that applying the geostrophic reconstruction procedure improves the accuracy for both the second- and third-order schemes. However, in the third-order case, the improvement is less pronounced. These results are consistent with those reported in \cite{Chertock2017}, where our second-order scheme exhibits slightly better performance in preserving the stationary vortex structure.

\subsection{Numerical order test}
The objective of this test is to empirically verify the numerical order of convergence for the two main numerical schemes, Geos-AWB2 and Geos-AWB3. The bottom topography and the initial water free surface are represented by a Gaussian hump defined as follows:
\begin{equation} \label{test_5_2_h}
    H(\theta,\varphi)=1-0.5e^{-\frac{\theta^2+\varphi^2}{100}}, \quad h(\theta,\varphi,0)=1-0.5e^{-\frac{\theta^2+\varphi^2}{100}}+0.1e^{-\frac{\theta^2+\varphi^2}{50}}.
\end{equation}

%
\noindent and the initial velocities are given by:
\begin{equation} \label{test_5_2_vel}
    u_{\theta}(\theta,\varphi,0)=-0.1\varphi e^{-\frac{\theta^2+\varphi^2}{50}}, \quad u_{\varphi}(\theta,\varphi,0)=0.1\theta e^{-\frac{\theta^2+\varphi^2}{50}}.
\end{equation}

\noindent The computational domain in decimal degrees is $[-180^{\circ},180^{\circ}]\times[-88^{\circ},88^{\circ}]$, where the singularities at the poles have been avoided. 
Boundary conditions are set as wall boundaries on the upper and lower edges, while periodic conditions are applied to the right and left limits. 
Five uniform structured grids with resolutions $\Delta \theta = \Delta \varphi = 8^{\circ}/2^l$, $l=0,\dots,4$ are considered for a sphere of radius $R=10,000$\,m. 
The simulation time is set to $T=600$\,s, with the CFL parameter fixed at 0.5. 
Since an exact solution is not available, a reference solution is computed on a finer grid with resolution $\Delta \theta = \Delta \varphi = 0.125^{\circ}$. 
Tables \ref{tablaOrden2} and \ref{tablaOrden3} present the relative errors of the conserved quantities in $L^1$-norm for the two numerical schemes at time 600\,s. 
As expected, second- and third-order accuracy is achieved.

\begin{table}[h!]
\begin{center}
\begin{tabular}{|c c c c c c c|} 

 \hline
 $\Delta \textbf{x}$ & Error $h_{\sigma}$ & Order & Error $Q_{\theta}$ & Order & Error $Q_{\varphi}$ & Order\\ [0.5ex] 
 \hline\hline
 $8^{\circ}$ & 1.05E-3 & --- & 3.38E-1 & --- & 2.90E-1 & --- \\ 
 \hline
 $4^{\circ}$ & 3.08E-4 & 1.79 & 1.22E-1 & 1.46 & 1.20E-1 & 1.27 \\
 \hline
 $2^{\circ}$ & 8.33E-5 & 1.89 & 3.05E-2 & 2.01 & 3.01E-2 & 1.99 \\
 \hline
 $1^{\circ}$ & 2.09E-5 & 1.99 & 7.56E-3 & 2.01 & 7.48E-3 & 2.01 \\
 \hline
 $0.5^{\circ}$ & 5.01E-6 & 2.06 & 1.84E-3 & 2.04 & 1.82E-3 & 2.03 \\
 \hline

\end{tabular}
\caption{Error in $L^1$-norm and order of convergence for Geos-AWB2. \label{tablaOrden2}}
\end{center}
\end{table}
\begin{table}[h!]
\begin{center}
\begin{tabular}{|c c c c c c c|} 

 \hline
 $\Delta \textbf{x}$ & Error $h_{\sigma}$ & Order & Error $Q_{\theta}$ & Order & Error $Q_{\varphi}$ & Order\\ [0.5ex] 
 \hline\hline
 $8^{\circ}$ & 3.07E-4 & --- & 4.04E-1 & --- & 3.96E-1 & --- \\ 
 \hline
 $4^{\circ}$ & 1.18E-4 & 1.37 & 1.70E-1 & 1.24 & 1.71E-1 & 1.21 \\
 \hline
 $2^{\circ}$ & 2.35E-5 & 2.33 & 3.70E-2 & 2.20 & 3.88E-2 & 2.14 \\
 \hline
 $1^{\circ}$ & 3.29E-6 & 2.84 & 5.33E-3 & 2.79 & 5.60E-3 & 2.79 \\
 \hline
 $0.5^{\circ}$ & 4.14E-7 & 2.99 & 6.74E-4 & 2.98 & 7.09E-4 & 2.98 \\
 \hline

\end{tabular}
\caption{Error in $L^1$-norm and order of convergence for Geos-AWB3. \label{tablaOrden3}}
\end{center}
\end{table}




\subsection{Test for the spherical geostrophic equilibrium}
This test, extracted from \cite{Galewsky2004}, consists on a basic zonal flow that represents a mid-latitude planetary jet. 
This test aims to evaluate the performance of the numerical scheme in the presence of a stationary solution of the complete system (\ref{2DRotatingSWESphere}) with a uniform atmospheric pressure field. The stationary solution is defined from the velocity field, and the initial fluid height layer is estimated from the balance equation via numerical integration. 
The latitudinal velocity component is set to zero $u_{\varphi}=0$, and the zonal velocity component $u_{\theta}$ is defined as a function of latitude. In radians, it is expressed as:
\begin{equation} \label{test_5_3_ux}
    u_{\theta}(\varphi) = \frac{u_{\mbox{\tiny max}}}{e_n} \mbox{exp}\bigg({\frac{1}{(\varphi - \varphi_0)(\varphi - \varphi_1)}}\bigg)\mathcal{X}_{(\varphi_0,\varphi_1)}(\varphi),
\end{equation}
\noindent being $\mathcal{X}_{(\varphi_0,\varphi_1)}(\varphi)$ the characteristic function in the interval $(\varphi_0,\varphi_1)$.
This expression sets the jet to be contained within a strip between latitudes $\varphi_0=\frac{\pi}{7}$ and $\varphi_1=\frac{5\pi}{14}$ with longitudinal boundaries at $-\pi$ and $\pi$.
The maximum zonal velocity $u_{\mbox{\tiny max}}=80\,$m/s occurs at the central latitude of the strip domain $[-\pi,\pi]\times[\varphi_0, \varphi_1]$ and smoothly decreases to zero outside the region defined by $[\varphi_0, \varphi_1]$.
The non-dimensional parameter $e_n=\frac{-4}{(\varphi_1-\varphi_0)^2}$ serves to normalize the velocity profile.
This function was introduced by Galewsky et al. \cite{Galewsky2004} to solve some difficulties in the numerical test suite proposed by Williamson et al. \cite{Williamson1992}. This definition of $u_\theta$ has some good properties such as being infinitely differentiable and compactly supported within the strip. The initial water height can be derived assuming that $h$ only depends upon latitude and working on the third equation in (\ref{2DRotatingSWESphere}), as the other two are trivially satisfied (see supplementary material (SM2) for the complete derivation). We set $h_0=10100\,$m to be the global mean height. Figure \ref{test_5_3_initial_figures} depicts these initial conditions.
%
\begin{figure}[h]
    \centering
    \includegraphics[height=0.4\textwidth]{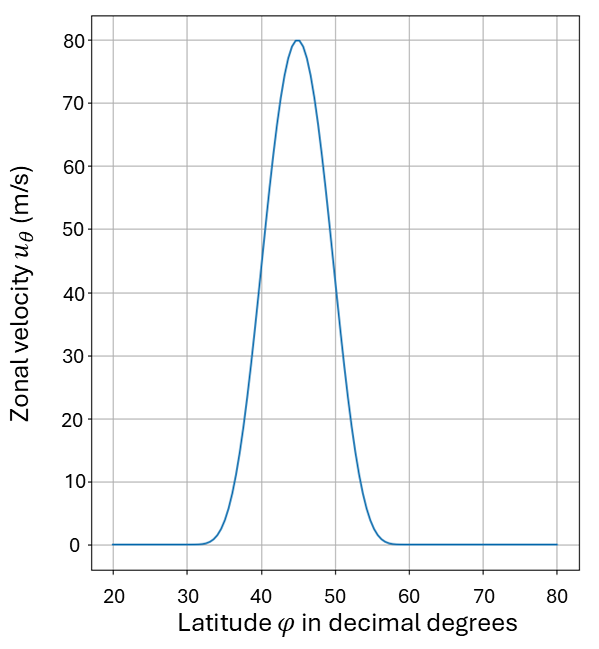}
    \includegraphics[height=0.4\textwidth]{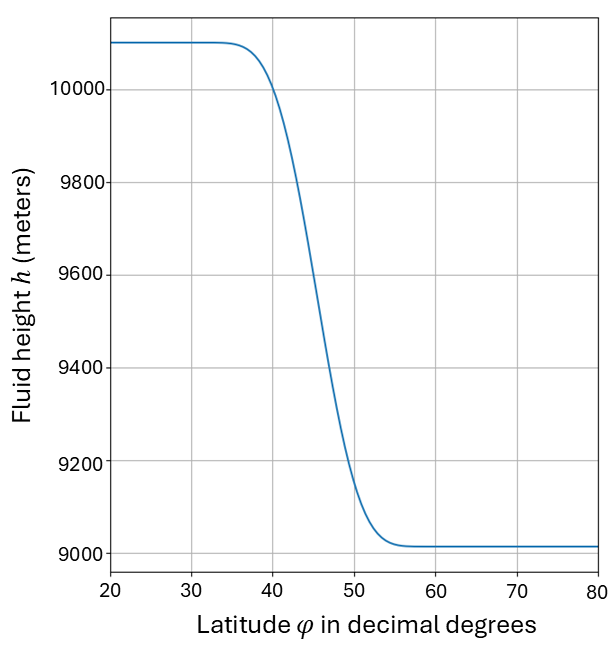}
    \caption{Initial conditions for the zonal flow test. Left: zonal velocity component, $u_{\theta}$, as defined in (\ref{test_5_3_ux}). Right: total fluid height, $h$, obtained from (SM2.2) using a quadrature rule. \label{test_5_3_initial_figures}}
\end{figure}

The computational domain is defined as $D=[-180^{\circ},180^{\circ}] \times [20^{\circ}, 80^{\circ}]$ to include the strip $[-\pi,\pi]\times[\varphi_0, \varphi_1]$. 
The spatial resolution is set to $\Delta \theta=\Delta \varphi=0.5^{\circ}$. 
Boundary conditions are configured as open at the upper and lower boundaries, while periodic conditions are applied at the right and left boundaries.  
The system is integrated over a period of five days ($T=432000\,$s), with results saved every 600\,s, and CFL number is set to 0.4. 
Figure \ref{errors_test_galewsky} shows the performance of the four numerical schemes by evaluating the evolution of the relative errors, computed in the $L^1$ norm relative to the exact solution. 
These results highlight the relevance of performing the geostrophic reconstruction, as evidenced by the significant difference in error evolution between Geos-NAWB2 and Geos-AWB2. The third-order non-well-balanced scheme, Geos-NAWB3, achieves better accuracy than its second-order well-balanced counterpart, Geos-AWB2, albeit at a higher computational cost. Finally, it can be observed that the third-order well-balanced numerical scheme Geos-AWB3, demonstrates exceptional stability, maintaining a nearly constant error over time.

{
Interestingly, the evolution of the numerical errors in this test shows significantly better behavior compared to those observed in the stationary vortex case (Figure~\ref{errors_test_chertock}). As the current test is characterized by a zero latitudinal velocity, and the simulation is carried out on a structured grid, whose orientation aligns with the velocity field, it is reasonable that the asymptotic well-balanced schemes yield better numerical error evolution than in the vortex test. In the latter, the exact stationary solution involves a non-zero velocity field in both directions, which increases the risk of small numerical error leakage across both components.
}
\begin{figure}[h]
    \centering
    \includegraphics[height=0.4\textwidth]{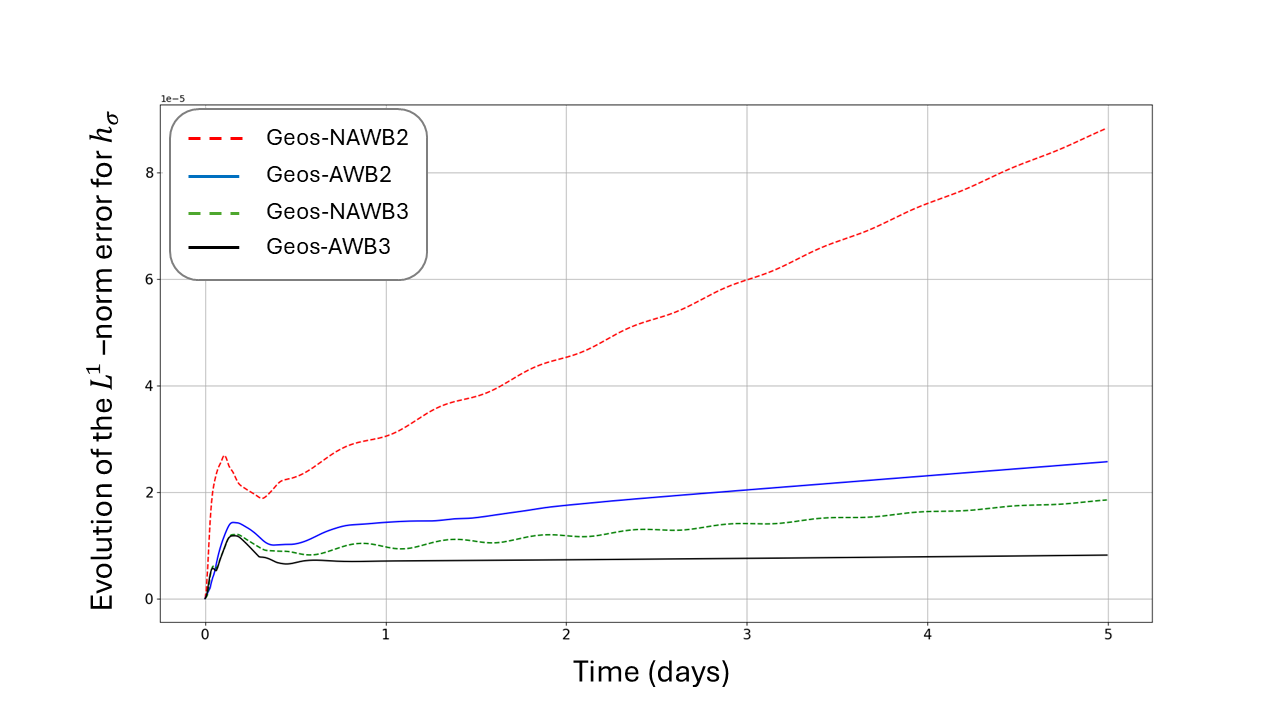}

    \caption{Relative errors for the evolution of $h_{\sigma}$ in $L^{1}$-norm for the zonal jet flow test. Both asymptotic well-balanced schemes Geos-AWB2, Geos-AWB3 evidence better results in presence of a stationary solution. \label{errors_test_galewsky}}.
\end{figure}

\subsection{Test for a perturbation of the spherical geostrophic equilibrium}
The goal of this test is to evaluate the ability of the numerical scheme to deal with complex structures across different time scales. It builds directly on the previous example, where the steady zonal flow is now perturbed by introducing a localized hump in the initial fluid height. This perturbation of the initial fluid height leads to complex dynamics at different time scales, where fast inertia gravity waves are developed within the first hours of the fluid simulation, giving way in the long term to barotropic instabilities dominated by vorticity dynamics. {Qualitatively, the results obtained with the Geos-AWB2 and Geos-AWB3 schemes exhibit a high degree of similarity. If computational cost is a priority, the second-order scheme is sufficient to capture the essential dynamics of the test. However, the third-order scheme offers sharper results due to its reduced numerical diffusion, albeit at the cost of increased computational effort. For this reason, we present here only the results obtained with the Geos-AWB3 scheme.}
In any case, the primary objective of this test is to demonstrate the capability of the designed numerical schemes to replicate these flow dynamics, producing results analogous to those reported in \cite{Galewsky2004}.

\begin{figure}
    \centering
    \includegraphics[height=0.4\textwidth]{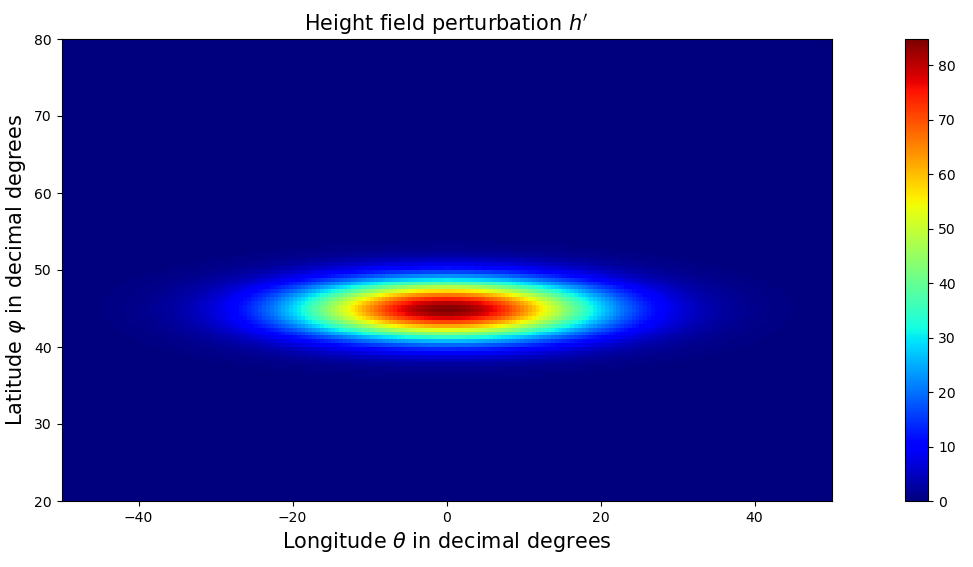}

    \caption{Perturbation of the height field $h'$ as defined in (\ref{test_perturbation_hprima_function}).  \label{test_perturbation_hprima_figure}}.
\end{figure}

\noindent The local hump is defined as
\begin{equation} \label{test_perturbation_hprima_function}
    h'(\theta, \varphi)=\Tilde{h}\, \mbox{cos}\,\varphi \, e^{-(\frac{\theta}{\alpha})^2}e^{-\big[\frac{(\varphi_2-\varphi)}{\beta}\big]^2},
\end{equation}
\noindent where $\varphi_2=\pi /4$, $\alpha=1/3$, $\beta=1/15$, and $\Tilde{h}=120\,$m. Figure \ref{test_perturbation_hprima_figure} shows the 2D spatial distribution of the perturbation height field. 
The initial velocity field is the same as in the previous test, while the initial fluid height is initialized as $\hat{h}=h+h^{'}$, where $h$ is the unperturbed height field previously computed through the integral equation (SM2.2). 
The plot on the left in Figure~\ref{test_perturbation_hdiff} presents several snapshots of the evolution of the simulated perturbation, showing the difference between the instantaneous height $\hat{h}$ and the balanced, unperturbed initial height $h$ as shown in Figure~\ref{test_5_3_initial_figures} (right). As expected in a Rossby adjustment, inertia-gravity waves radiate outward from the center of the initial perturbation.
The plot on the right in Figure \ref{test_perturbation_hdiff} shows the evolution of the relative potential vorticity, defined as $PV=\frac{w+f}{h}$. Over time, the vorticity field begins to exhibit signs of instability, ultimately evolving into a barotropic instability. By the fourth day of simulation, the instability becomes pronounced, manifesting as tightly wound vortices with steep gradients that progressively drift eastward by $t = 5$ days.

\begin{figure}[H]
    \centering
    \includegraphics[height=0.45\textwidth]{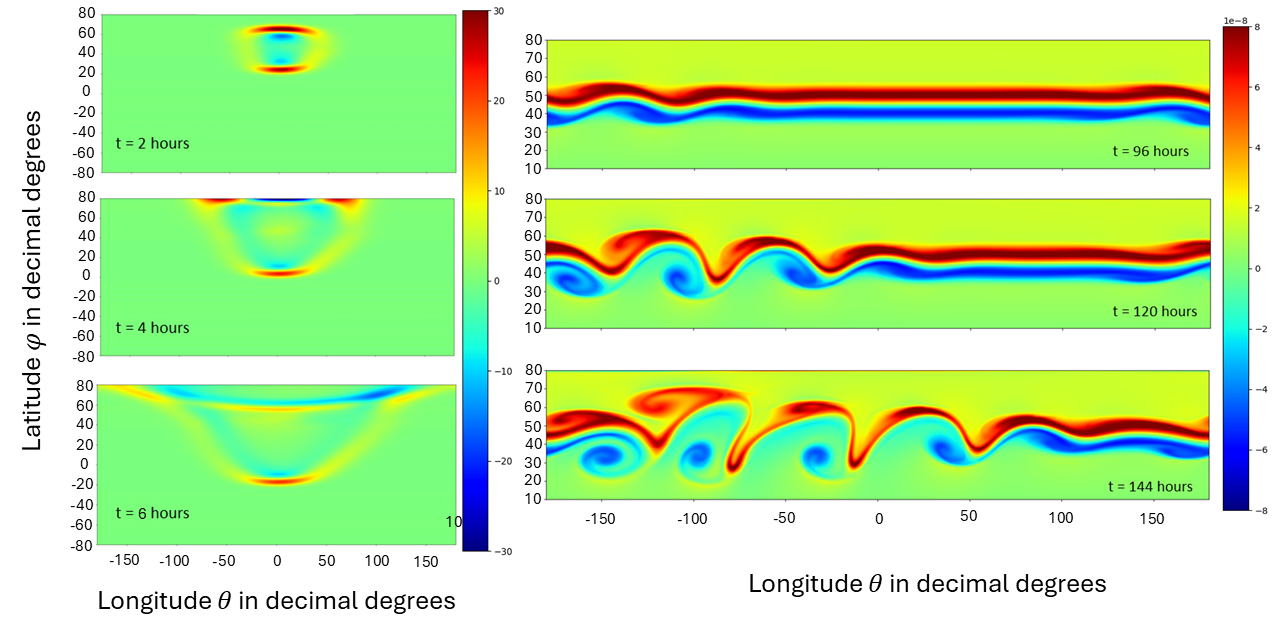}
    \caption{Left: snapshots of the height perturbation computed using the Geos-AWB3 numerical scheme at times 2h (top), 4h (middle) and 6h (bottom). These fast inertia gravity waves predominate during the initial moments of the flow simulation. Right: snapshots of the relative potential vorticity field computed using the Geos-AWB3 numerical scheme are shown for times 96\,h (top), 120\,h (middle) and 144\,h (bottom). The initially unstable vorticity evolves into multiple eddies (middle, bottom), forming very tight gradients and moving eastward in a characteristic way for atmospheric flows.\label{test_perturbation_hdiff}}.
\end{figure}



\subsection{Test for an observed meteotsunami (NTHMP)}
The last numerical test focuses on reproducing a real meteotsunami event that occurred in the Gulf of Mexico (GOM) in 2010. This test is based primarily on studies initiated by NOAA in 2020 as part of the National Tsunami Hazard Mitigation Program (NTHMP), with the final report published in 2022 \cite{NTHMP2022}. 
These studies are included within the framework of \textit{Meteotsunami Guidelines and Best Practices} for validating meteotsunamis numerical simulations. 
The 2010 event was driven by a dry season storm, during which low-pressure systems typically move eastward, and squall lines form ahead of cold fronts \cite{Olabarrieta2017}. 
In this case, a squall line developed off the Florida coast near Panama City and moved southeast along the isobar of the southwestern Florida shelf.
This triggered a long water wave with a height of approximately $1\,\text{m}$, which, amplified by Proudman resonance, caused inundation at Clearwater Beach, Florida.

To reproduce the event, synthetic and measured data are used.  
Bathymetry data has been downloaded from the global relief model ETOPO1 (\url{https://maps.ngdc.noaa.gov/viewers/wcs-client/
}) at a resolution of $15$ arc-seconds ($\approx450\,$m). 
Figure \ref{pressureProfile} (left) shows the area covered by this topobathymetry. 
An idealized analytic 2D pressure profile is defined and moved along the trajectory from point $A=(-85.107^{\circ},\, 30.7470^{\circ})$ to point $B=(-81.562^{\circ},23.4480^{\circ})$ at a constant speed of $U=20\,$m/s.
This motion takes approximately 13.5\,h to complete the path. 
The analytical expression of the atmospheric disturbance in decimal degrees units is given by:
\begin{equation} \label{test_5_5_analyticPressure}
P(x,y)=
\begin{cases}
    -A_{L}x e^{-(\frac{y}{C})^2-(\frac{x}{B_L})^2}, \quad x<0, \\
    -A_{R}x e^{-(\frac{y}{C})^2-(\frac{x}{B_R})^2}, \quad x>0,  \\
\end{cases}
\end{equation}
\noindent where the parameters are $A_L=37.61$, $B_L=0.31$, $A_R=4.665$, $B_R=0.5$, $C=0.5$. 
This pressure profile exhibits a maximum of $5\,$hPa and a minimum of $-1\,$hPa (see the right plot in Figure \ref{pressureProfile}). 
To better mimic the atmospheric disturbance pattern, a curvature is incorporated into the pressure profile using a stereographic projection.
This projection maps the disturbance onto a circular section defined by two concentric circumferences with radius $8^{\circ}$ and $11^{\circ}$. 
Finally, the resulting pressure is centered at the starting point $A$ and rotated to align with the trajectory toward point $B$. Left plot in Figure \ref{pressureProfile} shows the initial configuration for the simulation.

\begin{figure}
    \centering
    \includegraphics[height=0.40\textwidth]{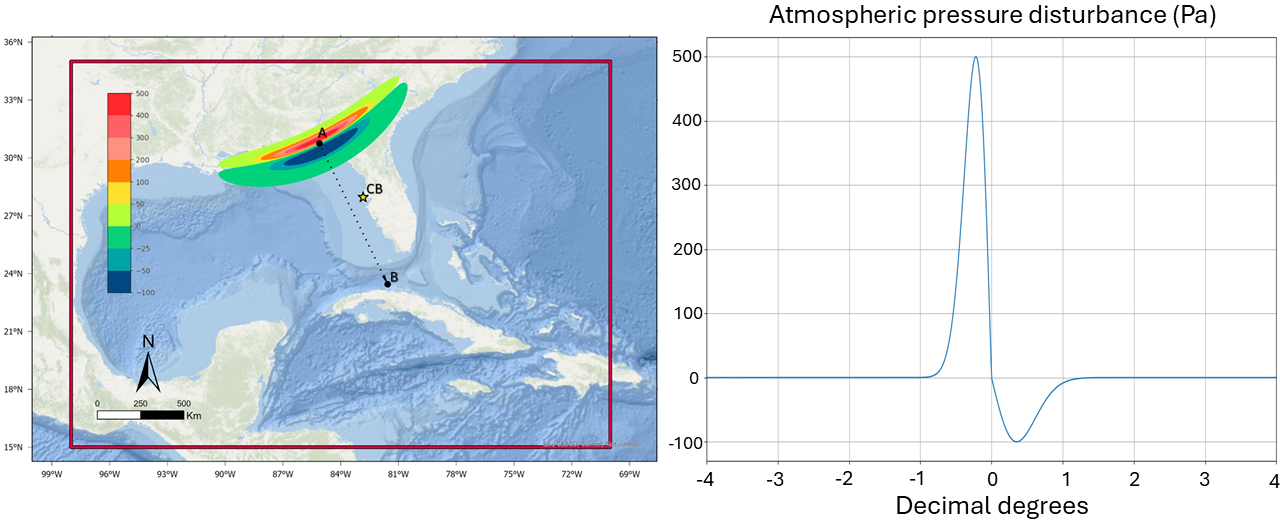}
    \caption{Left: Region of interest and initial condition setup for the meteotsunami simulation test case. Red rectangle determines the area covered by the global relief model ETOPO1 downloaded from the NOAA database at 15 arc-seconds spatial resolution. Pressures moves from point A to point B. Dashed line represents the traveling path. Yellow star marker (CB) is the location of Clearwater Beach. Right: Cross section at $y=0$ of the analytic pressure (\ref{test_5_5_analyticPressure}). It reaches a maximum of $5\,$hPa and a minimum of $-1\,$hPa.  \label{pressureProfile}}
\end{figure}



The simulation is performed using the second-order numerical scheme (Geos-AWB2) with constant Manning bottom friction coefficient of $\tau=0.025\,\mbox{sm}^{1/3}$ in the rectangular domain $[-98^{\circ},-70^{\circ}]\times[15^{\circ},35^{\circ}]$ using the WGS84 geographic reference system (epsg: 4326). 
The computational grid consists of $6720\times4800 = 32256000$ volumes, and the simulation time is set to $20\,$h. 
A wet and dry technique has been applied to the reconstruction operators following the method described in \cite{Gallardo2008}. 
Figure \ref{2dSim_0} shows snapshots of the simulation within a clipped domain centered around the coast of Florida.
\begin{figure}
    \centering
    \includegraphics[height=0.5\textwidth]{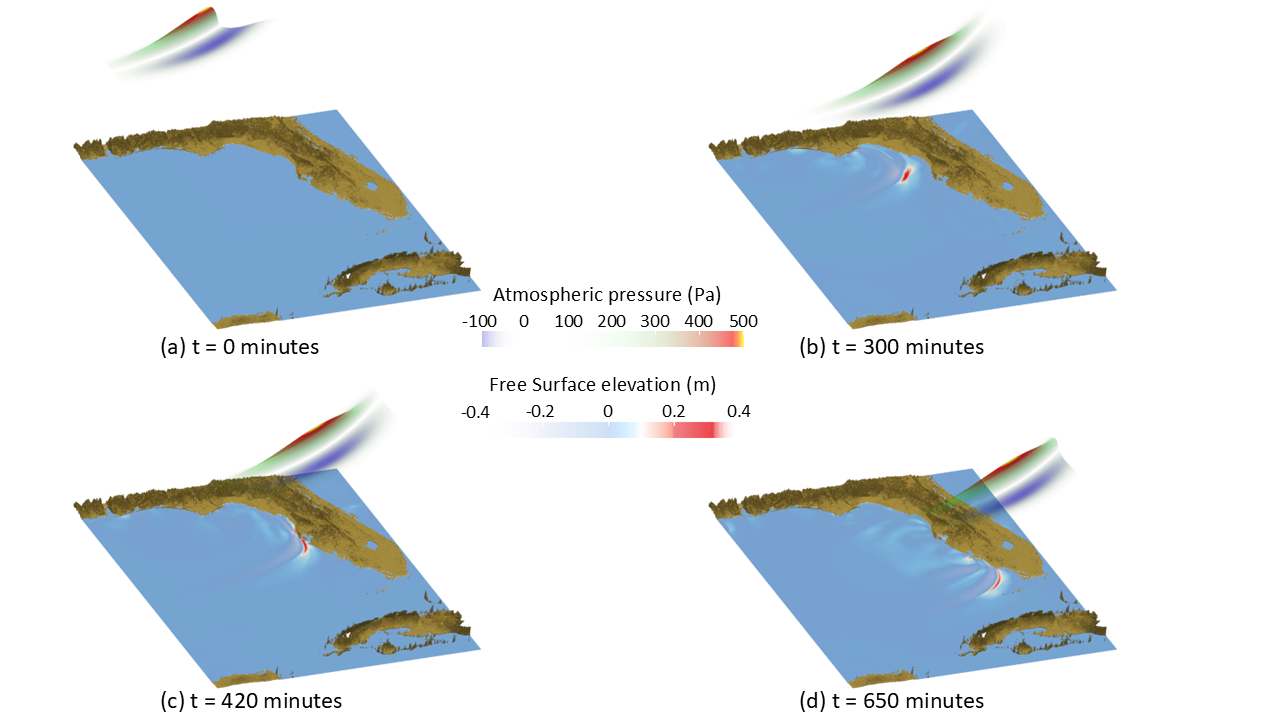}
    \caption{Snapshots of the meteotsunami simulation for the 2010 event in the GOM. Both pressure disturbance and water free surface are represented at different times.  \label{2dSim_0}}
\end{figure}



Finally, Figure \ref{mtHySEA_ts} presents a time series for the Clearwater Beach location at $(-82.83167^{\circ}, 27.97834^{\circ})$, where tide gauge measures were available. 
An overall good match is achieved, successfully capturing the recorded extreme peak values.
The arrival time of the largest wave, around 6-7 hours into the simulation, is quite accurately predicted. 
These results are consistent and in good agreement with those reported in the NOAA report \cite{NTHMP2022}. 
However, further tuning of the constants defining the pressure disturbance field (\ref{test_5_5_analyticPressure}) could potentially enhance the accuracy of the predicted arrival time.

\begin{figure}[H]
    \centering
    \includegraphics[height=0.45\textwidth]{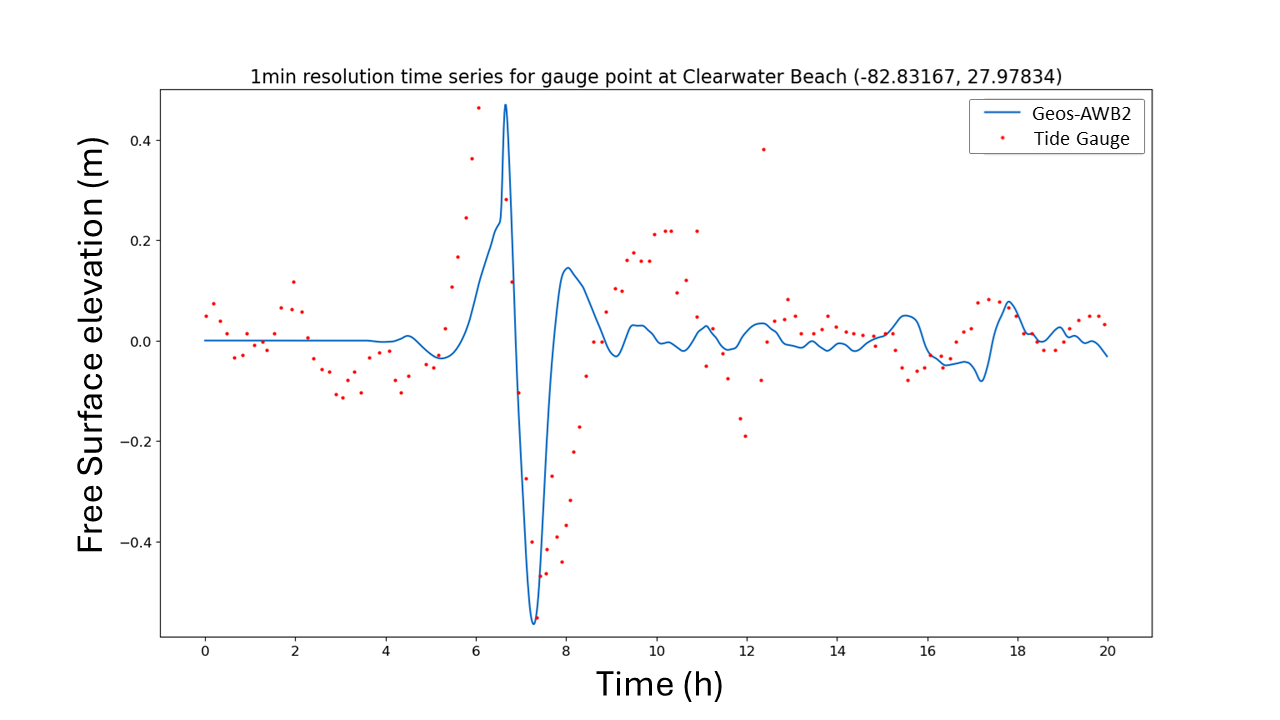}
    \caption{Time series retrieved at Clearwater Beach. Red dots are real data gathered from a tide gauge.  \label{mtHySEA_ts}}
\end{figure}

\section{Conclusions}   \label{sec:conclusions}
In this work, two high-order finite volume asymptotic well-balanced numerical schemes for the two-dimensional rotating NLSWE have been presented. Through the use of reconstruction operators, high-order spatial accuracy was achieved. Furthermore, well-balanced local linear and quadratic reconstructions were implemented for the geostrophic equilibrium in spherical coordinates. These reconstructions demonstrate the asymptotic well-balanced behavior of the numerical approximation for low Rossby and Froude numbers.

A GPU parallel implementation has been performed to reduce the computational time. Several numerical tests have been conducted to assess the numerical order of convergence and to highlight the importance of the geostrophic reconstruction in accurately modeling stationary states near the geostrophic equilibrium.

Finally, a simulation was presented that illustrates the capability of the numerical scheme to accurately reproduce meteotsunamis over realistic bathymetries. Future work will focus on incorporating a nested grid system to enhance meteotsunami predictions along continental shelves and coastlines. Given the wide range of spatial scales involved in such phenomena, it is crucial to capture the intricate details of nearshore geometry to provide reliable inundation forecasts.



\section*{Acknowledgments}
We would like to acknowledge Professor Carlos Parés for the ideas and corrections contributed during the development of this work.

\bibliographystyle{siamplain}

\end{document}